\documentclass[12pt,a4paper]{article}
\usepackage{CJK}
\usepackage{amssymb}
\usepackage[top=1in, bottom=1in, left=0.5in, right=0.5in]{geometry}
\usepackage{indentfirst}
\usepackage{hyperref}
\usepackage[numbers,sort&compress]{natbib}
\usepackage{color}

\begin{document}

\begin{titlepage}

\vspace{10mm}
\begin{center}
{\Large\bf One-Loop Renormalizable Wess-Zumino Model on Bosonic-Fermionic Noncommutative Superspace}
\vspace{16mm}

{\large Yan-Gang Miao\footnote{E-mail address: miaoyg@nankai.edu.cn} and Xu-Dong Wang\footnote{E-mail address: c\_d\_wang@mail.nankai.edu.cn}

\vspace{6mm}
{\normalsize \em School of Physics, Nankai University, Tianjin 300071, China}}

\end{center}

\vspace{10mm}
\centerline{{\bf{Abstract}}}
\vspace{6mm}
We construct a deformed Wess-Zumino model on the noncommutative superspace where the Bosonic and Fermionic coordinates are no longer commutative with each other. Using the background field method, we calculate the primary one-loop effective action based on the deformed action. By comparing the two actions, we find that the deformed Wess-Zumino model is not renormalizable. To obtain a renormalizable model, we combine the primary one-loop effective action with the deformed action, and then calculate the secondary one-loop effective action based on the combined action. After repeating this process to the third time, we finally give the one-loop renormalizable action up to the second order of Bosonic-Fermionic noncommutative parameters by using our specific techniques of calculation. 

\vskip 20pt
\noindent
{\bf PACS Number(s)}: 12.60.Jv, 11.30.Pb, 11.10.Nx, 11.10.Gh


\end{titlepage}

\newpage

\tableofcontents

\newpage

\section{Introduction}

There are a lot of studies that generalize the ordinary spacetime to a noncommutative (NC) spacetime and provide interesting physical models~\cite{Seiberg:1999vs, Douglas:2001ba,Szabo:2001kg}. Because these studies rely only on a formal extension without experiments, there are many possible choices to construct NC spacetimes. A different kind of NC spacetimes usually gives different results that should generally not contradict with the very basic physical principle. 

As an example, we take the canonical NC spacetime defined by
\begin{equation} 
\label{} 
\left[x^{\mu }, x^{\nu }\right]=i \theta ^{\mu \nu }, \qquad \mu ,\nu =0,{\cdots},D-1,
\end{equation} 
where $\theta ^{\mu \nu }$ is a real antisymmetric constant tensor.
In ref.~\cite{Gomis:2000zz} it is verified  that a different choice of NC parameter $\theta ^{\mu \nu }$ changes the property of the model defined on it.
When  $\theta ^{0i}=0$ the model is unitary, but non-unitary while $\theta ^{0i}\neq 0$.  
The former can be obtained from string theory while the latter can not. In consequence, 
the fact that the model defined on the canonical NC spacetime with $\theta ^{0i}=0$ is unitary results from the fact that string theory is unitary.

The above explanation is argued by,
for instance, refs.~\cite{Doplicher:1994tu,Bahns:2002vm} where it is claimed that the breakdown of unitarity of the model defined on the canonical NC spacetime with  $\theta ^{0i}\neq 0$ may have another origin, i.e., a wrong method to define the field theory on the NC spacetime may be utilized. The unitarity of the model can be restored if a different definition of field theory on the NC spacetime is given.
Nonetheless, we note that it is not a generally accepted method to modify the definition of field theory on NC spacetimes.

Moreover, the ordinary spacetime can be extended to include Fermionic coordinates that are anticommutative to each other, which gives rise to a superspace~\cite{Nilles:1983ge,Wess:1992cp,Gates:1983nr}. Therefore, one interesting generalization is a non-anticommutative (NAC) superspace~\cite{Ferrara:2000mm, Klemm:2001yu, Ooguri:2003qp, Ooguri:2003tt, Seiberg:2003yz, Berkovits:2003kj, Ferrara:2003xy}. For example, in ref.~\cite{Seiberg:2003yz} the ${\cal N}=1$ superspace is generalized to a NAC formulation. 
Based on the anticommutative relations of Fermionic coordinates, the algebraic relations of the other coordinates can be determined. Such a deformation from the anticommutative case to the non-anticommutative case is realized by using the star($\star$)-product associated with the NAC superspace. As a result, the Wess-Zumino model and the super Yang-Mills theory can consistently be defined on the NAC superspace.

Ref.~\cite{Seiberg:2003yz}  has inspired a lot of researches on the construction of models on the NAC superspace.
In ref.~\cite{Britto:2003aj} the renormalization property of the Wess-Zumino model on the NAC superspace is studied and new (non)renormalization theorems are proved.
In addition, two global $U(1)$ symmetries are found for the NAC Wess-Zumino model. 
In ref.~\cite{Grisaru:2003fd} the NAC Wess-Zumino model is investigated by calculating the 1PI effective action. At the one-loop order, new terms appear, which makes the model not renormalizable. To solve this problem, the new terms are added to the NAC Wess-Zumino action. With these new terms, the NAC Wess-Zumino model is proved to be renormalizable up to two loops. 
This work raises the problem that whether the NAC Wess-Zumino model can be renormalizable to higher order loop expansion. 
To answer this question, the dimensional analysis method is applied and the NAC Wess-Zumino model
is thus proved~\cite{Britto:2003kg} to be renormalizable to all loop orders in terms of the component form.   
Furthermore, the renormalizability can be retrieved~\cite{Romagnoni:2003xt} in terms of the superfield form. It should be emphasized that the key point of the proofs in refs.~\cite{Britto:2003kg,Romagnoni:2003xt} is the existence of the two global $U(1)$ symmetries discovered in ref.~\cite{Britto:2003aj}. 

Beyond an individual NC spacetime and an individual NAC superspace, the combination of a NC spacetime and a NAC superspace brings further interest, where the Bosonic-Fermionic noncommutative (BFNC) superspace has been considered by the argument of string theory~\cite{de Boer:2003dn}. 
On  the BFNC superspace the Bosonic coordinates are no longer commutative with the Fermionic coordinates. 
A natural motivation is thus to study physical consequences of the BFNC superspace for some interesting  models. To our knowledge, the properest choice is the Wess-Zumino model and the most appealing physical consequence among the many properties related to the model is renormalizability.

In this paper we construct the deformed Wess-Zumino model on the BFNC superspace and indeed work out its one-loop renormalizable action up to the second order of BFNC parameters.  
 The paper is organized as follows.  In section~\ref{Supersymmetry_and_Wess-Zumino_model} we review the ${\cal N}=1$ supersymmetry and the Wess-Zumino model on the ordinary superspace in order to make this paper self-contained. 
In section~\ref{Deformed_Action_From_star_Product} we derive the deformed Wess-Zumino action through substituting the ordinary product by the BFNC star($\star$)-product in the action of the (ordinary) Wess-Zumino model. 
For the purpose of searching a renormalizable Wess-Zumino action on the BFNC superspace in light of the background field method, we briefly review this method  in  section~\ref{Background_Field_Method}. 
In terms of our specific techniques of calculation, such as establishing the supersymmetry invariant subsets and their bases, we give the one-loop renormalizable action up to the second order of BFNC parameters in section~\ref{Main_Result}. 
At last we present our conclusion and outlook in section~\ref{Conclusion_and_Outlook}. Incidentally,  some useful formulae and the final long results  are put into Appendices A and B.

\section{${\cal N}=1$ Supersymmetry and Wess-Zumino Model}
\label{Supersymmetry_and_Wess-Zumino_model}

In this section we briefly review the supersymmetry~\cite{Nilles:1983ge} and the Wess-Zumino model on the ordinary superspace for the sake of making our discussions self-contained. We use the conventions given by Wess and Bagger~\cite{Wess:1992cp}.

The ${\cal N}=1$ superspace is represented by coordinates $x^k$, $\theta ^{\alpha }$, and $\bar{\theta }^{\dot{\alpha }}$, where $k=0, 1, 2, 3$, and $\alpha =1, 2$. 
$x^k$ is a commutative number, while $\theta ^{\alpha }$ is an anticommutative number.

In the chiral coordinates $y^k$, $\theta ^{\alpha }$, and $\bar{\theta }^{\dot{\alpha }}$, 
where 
\begin{equation}
\label{chiral coordinates}
y^k\equiv x^k+i \sigma ^k{}_{\alpha  \dot{\beta }} \theta ^{\alpha } \bar{\theta }^{\dot{\beta }},
\end{equation}

$Q$ and $D$ are defined by
\begin{eqnarray}
\label{chiral_representation}
Q_{\alpha }&\equiv&\frac{\partial }{\partial \theta ^{\alpha }},\qquad 
\bar{Q}_{\dot{\alpha }}\equiv-\frac{\partial }{\partial \bar{\theta }^{\dot{\alpha }}}+2 i \theta ^{\beta } \sigma ^k{}_{\beta  \dot{\alpha }} \frac{\partial }{\partial y^k},\nonumber\\
D_{\alpha }&\equiv&\frac{\partial }{\partial \theta ^{\alpha }}+2 i \sigma ^k{}_{\alpha  \dot{\beta }} \bar{\theta }^{\dot{\beta }} \frac{\partial }{\partial y^k},\qquad 
\bar{D}_{\dot{\alpha }}\equiv-\frac{\partial }{\partial \bar{\theta }^{\dot{\alpha }}}.
\end{eqnarray}

The operators $Q$ and $D$ satisfy the following algebraic relations,
\begin{eqnarray}
\label{algebra_relation_for_generator}
\left\{Q_{\alpha },\bar{Q}_{\dot{\beta }}\right\}&=&2i \sigma ^k{}_{\alpha  \dot{\beta }}\partial_k, \qquad 
\left\{Q_{\alpha },Q_{\beta }\right\}=\left\{\bar{Q}_{\dot{\alpha }},\bar{Q}_{\dot{\beta }}\right\}=0,\nonumber\\
\left\{D_{\alpha },\bar{D}_{\dot{\beta }}\right\}&=&-2i \sigma ^k{}_{\alpha  \dot{\beta }}\partial_k, \qquad 
\left\{D_{\alpha },D_{\beta }\right\}=\left\{\bar{D}_{\dot{\alpha }},\bar{D}_{\dot{\beta }}\right\}=0,\nonumber\\
\left\{D_{\alpha },Q_{\beta }\right\} &=&\left\{D_{\alpha },\bar{Q}_{\dot{\beta }}\right\}=\left\{\bar{D}_{\dot{\alpha }},Q_{\beta }\right\}=\left\{\bar{D}_{\dot{\alpha }},\bar{Q}_{\dot{\beta }}\right\}=0,
\end{eqnarray}
where the bracket of two operators, say  $O_1$ and $O_2$, is defined as $\{O_1, O_2\} \equiv O_1 O_2+O_2 O_1$, 
and $\partial_k \equiv \frac{\partial }{\partial y^k}$.

The derivatives for anticommutative numbers take the forms,
\begin{equation}
\label{}
\frac{\partial }{\partial \theta ^{\alpha }}\theta ^{\beta }=\delta _{\alpha  \beta },\qquad \frac{\partial }{\partial \bar{\theta }^{\dot{\alpha }}}\bar{\theta }^{\dot{\beta }}=\delta _{\dot{\alpha }\dot{\beta }},
\end{equation}
where $\delta _{\alpha \beta }=\delta _{\dot{\alpha }\dot{\beta }}=1$ for $\alpha =\beta$ and ${\dot{\alpha }=\dot{\beta }}$, and 
$\delta _{\alpha \beta }=\delta _{\dot{\alpha }\dot{\beta }}=0$, for  $\alpha\neq \beta$ and ${\dot{\alpha }\neq \dot{\beta }}$.

\subsection{Chiral Superfield}

The chiral superfield $\Phi$ and the antichiral superfields $\Phi ^+$ satisfy the following conditions, respectively,
\begin{equation}
\label{chiralcondition}
\bar{D}_{\dot{\alpha }}\Phi =0, \qquad D_{\alpha }\Phi ^+=0.
\end{equation}
Here the chiral and antichiral superfields are defined by
\begin{eqnarray}
\Phi \equiv \Phi (y,\theta), \qquad \Phi ^+ \equiv \Phi ^+(y,\theta ,\bar{\theta }), \label{Phi symbol}
\end{eqnarray}
which can be expressed by component fields as follows,
\begin{eqnarray}
\label{chiral_superfield}
\Phi (y,\theta)&=&A(y)+\sqrt{2} \theta ^{\alpha } \psi _{\alpha }(y)+\theta ^{\alpha } \theta _{\alpha } F(y),\nonumber\\
\Phi ^+(y,\theta ,\bar{\theta })&=&A^*(y)+\sqrt{2} \bar{\theta }_{\dot{\alpha }} \bar{\psi }^{\dot{\alpha }}(y)+\bar{\theta }_{\dot{\alpha }} \bar{\theta }^{\dot{\alpha }} F^*(y)-2 i \sigma ^k{}_{\alpha  \dot{\beta }} \theta ^{\alpha } \bar{\theta }^{\dot{\beta }} \partial _kA^*(y)\nonumber\\
&&+i \sqrt{2} \sigma ^k{}_{\alpha  \dot{\beta }} \theta ^{\alpha } \bar{\theta }_{\dot{\zeta }} \bar{\theta }^{\dot{\zeta }} \partial _k\bar{\psi }^{\dot{\beta }}(y)+\eta ^{k l} \theta ^{\alpha } \theta _{\alpha } \bar{\theta }_{\dot{\beta }} \bar{\theta }^{\dot{\beta }} \partial _k\partial _lA^*(y),
\end{eqnarray}
where $A(y)$ and $F(y)$ are scalar fields, $\psi _{\alpha }(y)$ and $\bar{\psi }^{\dot{\alpha }}(y)$ are spinor fields.

Spinor indices are raised and lowered by using the antisymmetric symbols $\epsilon ^{\alpha \beta }$ and $\epsilon _{\alpha \beta }$, respectively,
\begin{equation}
\label{}
\psi ^{\alpha }\equiv\epsilon ^{\alpha \beta }\psi _{\beta }, \qquad \psi _{\alpha }\equiv\epsilon _{\alpha \beta }\psi ^{\beta },
\end{equation}
where  the antisymmetric symbols  satisfy the relation: $\epsilon _{\alpha \beta }\epsilon ^{\beta \gamma }=\delta _{\alpha \gamma }$.

In the following context, the abbreviations 
\begin{eqnarray}
\theta ^2\equiv \theta ^{\alpha }\theta _{\alpha }, \qquad \bar{\theta }^2\equiv \bar{\theta }_{\dot{\alpha }}\bar{\theta }^{\dot{\alpha }} \label{theta2}
\end{eqnarray}
are used.

\subsection{Supersymmetry Transformation}

The supersymmetry transformation of the superfield ${\cal F}$ is defined as
\begin{equation}
\label{supersymmetry_transformation_general_superfield}
\delta _{\xi }{\cal F}\equiv \left(\xi ^{\alpha }Q_{\alpha }-\bar{\xi }^{\dot{\beta }}\bar{Q}_{\dot{\beta }}\right){\cal F},
\end{equation}
where $\xi ^{\alpha }$ and $\bar{\xi }^{\dot{\beta }}$ are spinor parameters.

Applying the above transformation rule to the chiral superfield, one can obtain the transformations of component fields,
\begin{eqnarray}
\label{transformation of component fields}
\delta _{\xi }A&=&\sqrt{2} \xi ^{\alpha } \psi _{\alpha },\nonumber\\
\delta _{\xi }\psi _{\alpha }&=&\sqrt{2} \xi _{\alpha } F+ i\sqrt{2} \sigma ^k{}_{\alpha  \dot{\beta }} \bar{\xi }^{\dot{\beta }} \partial _kA,\nonumber\\
\delta _{\xi }F&=&i \sqrt{2} \sigma ^k{}_{\alpha  \dot{\beta }} \bar{\xi }^{\dot{\beta }} \partial _k\psi ^{\alpha }.
\end{eqnarray}
Note that in eq.~(\ref{supersymmetry_transformation_general_superfield}) both the superfield ${\cal F}$ and the operator $Q$ are  expressed in the chiral coordinates $y$, $\theta$, and $\bar{\theta }$ (see  eqs.~(\ref{chiral coordinates}) and (\ref{chiral_representation})).

\subsection{Wess-Zumino Model}

The action of the Wess-Zumino model is 
\begin{equation}
\label{WZ action}
{\cal S}_{\rm WZ}=\int d^4x\left\{
\Phi ^+\Phi |_{\theta ^2\bar{\theta }^2}+\frac{m}{2}\Phi  \Phi |_{\theta ^2}+\frac{g}{3}\Phi  \Phi  \Phi |_{\theta ^2}+\frac{m}{2}\Phi ^+ \Phi ^+|_{\bar{\theta }^2}+\frac{g}{3}\Phi ^+ \Phi ^+ \Phi ^+|_{\bar{\theta }^2}
\right\},
\end{equation}
where $\Phi$ and $\Phi^+$ are given by eq.~(\ref{chiral_superfield}), 
$\Phi ^+\Phi |_{\theta ^2\bar{\theta }^2}$ denotes the ${\theta ^2\bar{\theta }^2}$ component of $\Phi ^+\Phi$, 
and the other terms have the similar meaning.

The Wess-Zumino action eq.~(\ref{WZ action}) can be transformed to a total superspace integral,
\begin{eqnarray}
\label{WZ_superfield_form}
{\cal S}_{\rm WZ}&=&\int d^8z\left\{\Phi ^+\Phi -\frac{m}{8}\Phi  \left(\frac{D^2}{\square }\Phi \right)-\frac{m}{8}\Phi ^+ \left(\frac{\bar{D}^2}{\square }\Phi ^+\right)\right.\nonumber\\
&&\hspace{15mm}\left.-\frac{g}{12}\Phi  \Phi \left(\frac{D^2}{\square }\Phi \right)-\frac{g}{12}\Phi ^+ \Phi ^+ \left(\frac{\bar{D}^2}{\square }\Phi ^+\right)\right\},
\end{eqnarray}
where $d^8z \equiv d^4x d^2\theta d^2\bar{\theta}$, $D^2=\epsilon ^{\alpha \beta }D_{\beta }D_{\alpha }$, and $\bar{D}^2=\epsilon ^{\dot{\alpha }\dot{\beta }}\bar{D}_{\dot{\alpha }}\bar{D}_{\dot{\beta }}$.

\section{Deformed Action from BFNC $\star$-Product}
\label{Deformed_Action_From_star_Product}

In this section we at first give the $\star$-product of the BFNC superspace and its corresponding algebraic relations of coordinates, and then define the deformed Wess-Zumino model in the component form on the Euclidean BFNC superspace. For calculating effective actions in section~\ref{Main_Result} we further transform the deformed model into its superfield form.
In the following the Lorentz signatures are still used as pointed out by Seiberg~\cite{Seiberg:2003yz}.

\subsection{BFNC $\star$-Product}

The BFNC $\star$-product can be expressed in terms of the tensor algebraic notation which is frequently used in quantum group theory~\cite{Majid:1996kd},
\begin{equation}
\label{star product}
{\bf F} \star {\bf G}\equiv\mu \left\{ \exp \left[\frac{i}{2}\Lambda ^{k \alpha }\left(\frac{\partial }{\partial y^k}\otimes \frac{\partial }{\partial \theta ^{\alpha }}-\frac{\partial }{\partial \theta ^{\alpha }}\otimes \frac{\partial }{\partial y^k}\right)\right]\triangleright({\bf F}\otimes {\bf G}) \right\},
\end{equation}
where $\Lambda ^{k \alpha }$ denotes a BFNC parameter, $k=0,1,2,3$, and $\alpha=1,2$, which implies that there are eight independent BFNC parameter components in total.

For the notation in eq.~(\ref{star product}),
the tensor product is defined as 
\begin{equation}
\label{}
(A\otimes B)(C\otimes D)=(-1)^{|B| |C|}A C\otimes B D,
\end{equation}
where $|B|$ is the grade of $B$ that equals $1$  for a Bosonic element and $-1$ for a Fermionic one.  
The symbol $\triangleright$ represents an action of an operator on a function, and $\mu$ denotes the change from the tensor product $\otimes$ to an ordinary product, and ${\bf F}$ and ${\bf G}$ stand for any superfields.

The Taylor expansion of eq.~(\ref{star product}) takes the form,
\begin{eqnarray}
\label{star_product_expansion}
{\bf F}\star {\bf G}&=&{\bf F} {\bf G}-\frac{i}{2}\Lambda ^{k \alpha }\left(\partial _{\alpha }{\bf F}\right)\left(\partial _k{\bf G}\right)+(-1)^{|{\bf F}|}\frac{i}{2}\Lambda ^{k \alpha }\left(\partial _k{\bf F}\right)\left(\partial _{\alpha }{\bf G}\right) \nonumber \\
&&+\frac{1}{8}\Lambda ^{k \alpha }\Lambda ^{l \beta }\left(\partial _k\partial _l{\bf F}\right)\left(\partial _{\alpha }\partial _{\beta }{\bf G}\right)+\frac{1}{8}\Lambda ^{k \alpha }\Lambda ^{l \beta }\left(\partial _{\alpha }\partial _{\beta }{\bf F}\right)\left(\partial _k\partial _l{\bf G}\right) \nonumber \\
&&+(-1)^{|{\bf F}|}\frac{1}{4}\Lambda ^{k \alpha }\Lambda ^{l \beta }\left(\partial _{\beta }\partial _k{\bf F}\right)\left(\partial _{\alpha }\partial _l{\bf G}\right)\nonumber \\
&&-\frac{i}{16}\Lambda ^{k \alpha }\Lambda ^{l \beta }\Lambda ^{m \zeta }\left(\partial _{\alpha }\partial _l\partial _m{\bf F}\right)\left(\partial _{\beta }\partial _{\zeta }\partial _k{\bf G}\right)\nonumber \\
&&+(-1)^{|{\bf F}|}\frac{i}{16}\Lambda ^{k \alpha }\Lambda ^{l \beta }\Lambda ^{m \zeta }\left(\partial _{\alpha }\partial _{\beta }\partial _m{\bf F}\right)\left(\partial _{\zeta }\partial _k\partial _l{\bf G}\right)\nonumber \\
&&-\frac{1}{64}\Lambda ^{k \alpha }\Lambda ^{l \beta }\Lambda ^{m \zeta }\Lambda ^{n \iota }\left(\partial _{\alpha }\partial _{\zeta }\partial _l\partial _n{\bf F}\right)\left(\partial _{\beta }\partial _{\iota }\partial _k\partial _m{\bf G}\right),
\end{eqnarray}
where $\partial_k\equiv \frac{\partial }{\partial y^k}$, and $\partial _{\alpha }\equiv \frac{\partial }{\partial \theta ^{\alpha }}$.

As a simple application of the $\star$-product  eq.(\ref{star_product_expansion}), we calculate some $\star$-algebraic relations of coordinates,
\begin{eqnarray}
\label{algebra_relations}
\left\{y^k,\theta ^{\alpha }\right\}_{\star}&=&i \Lambda ^{k \alpha },
\qquad 
\left[x^k,y^l\right]_{\star}=-\sigma ^k{}_{\alpha  \dot{\beta }} \Lambda ^{l \alpha } \bar{\theta }^{\dot{\beta }},
\qquad 
\left[x^k,\theta ^{\alpha }\right]_{\star}=i \Lambda ^{k \alpha },
\nonumber\\ 
\left[x^k,x^l\right]_{\star}&=&\sigma ^l{}_{\alpha  \dot{\beta }} \Lambda ^{k \alpha } \bar{\theta }^{\dot{\beta }}-\sigma ^k{}_{\alpha  \dot{\beta }} \Lambda ^{l \alpha } \bar{\theta }^{\dot{\beta }},
\end{eqnarray}
where the algebraic relations in eq.~(\ref{algebra_relation_for_generator}) are not modified.

\subsection{Deformed Wess-Zumino Model}

To analyze  the effect of the BFNC superspace on the Wess-Zumino model, we replace the ordinary product in eq.~(\ref{WZ action}) by the $\star$-product defined by  eq.~(\ref{star product}) and give the deformed action,
\begin{equation}
\label{deformed WZ action}
{\cal S}_{\rm NC}\equiv
\int d^4x\left\{
\Phi ^+\star\Phi |_{\theta ^2\bar{\theta }^2}+\frac{m}{2}\Phi  \star \Phi |_{\theta ^2}+\frac{g}{3}\Phi  \star \Phi \star \Phi |_{\theta ^2}+\frac{m}{2}\Phi ^+ \star \Phi ^+|_{\bar{\theta }^2}+\frac{g}{3}\Phi ^+ \star  \Phi ^+ \star \Phi ^+|_{\bar{\theta }^2}
\right\},
\end{equation}
where the chiral superfield $\Phi$ and the antichiral superfield $\Phi^+$ are given in the chiral coordinates as  in eq.~(\ref{chiral_superfield}).

To obtain the explicit form of the deformed Wess-Zumino action, we  calculate the $\star$-product of the chiral and antichirl superfields,
\begin{eqnarray}
\label{star_of_chiral_field_component_form}
\int d^4x~ \Phi ^+\star \Phi |_{\theta ^2\bar{\theta }^2}&=&\int d^4x\left\{-i \sigma ^k{}_{\alpha  \dot{\beta }} \psi ^{\alpha } \partial _k\bar{\psi }^{\dot{\beta }}+A \square A^*+F F^*\right\},\nonumber\\
\int d^4x~ \Phi  \star \Phi  |_{\theta ^2}&=&\int d^4x\left\{2 A F-\psi ^{\alpha } \psi _{\alpha }\right\},\nonumber\\
\int d^4x~ \Phi ^+\star \Phi ^+|_{\bar{\theta }^2}&=&\int d^4x\left\{-\bar{\psi }_{\dot{\alpha }} \bar{\psi }^{\dot{\alpha }}+2 A^* F^*\right\},\nonumber\\
\int d^4x~ \Phi  \star \Phi  \star \Phi |_{\theta ^2}&=&\int d^4x\left\{-3 \psi ^{\alpha } \psi _{\alpha } A+3 A A F-\frac{3}{4} \Lambda ^{k l} \partial _k\partial _lA F F\right.\nonumber\\
&&\hspace{15mm}\left.+\frac{3}{2} \Lambda ^{k \alpha } \Lambda ^{l \beta } \partial _k\psi _{\beta } \partial _l\psi _{\alpha } F+\frac{1}{16} \Lambda ^{k l} \Lambda ^{n o} F \partial _k\partial _lF \partial _n\partial _oF\right\},\nonumber\\
\int d^4x~ \Phi ^+\star \Phi ^+\star \Phi ^+|_{\bar{\theta }^2}&=&\int d^4x\left\{-3 \bar{\psi }_{\dot{\alpha }} \bar{\psi }^{\dot{\alpha }} A^*+3 A^* A^* F^*-\frac{1}{2} \Lambda ^{k l} A^* \square A^* \partial _k\partial _lA^*\right.\nonumber\\
&&\hspace{15mm}+\frac{1}{2} \eta ^{k l} \Lambda ^{n o} A^* \partial _k\partial _nA^* \partial _l\partial _oA^*\nonumber\\
&&\hspace{15mm}\left.+\left(\sigma ^{k l}\right)_{\alpha }^{\beta } \Lambda ^{n \alpha } \Lambda ^o{}_{\beta } A^* \partial _k\partial _nA^* \partial _l\partial _oA^*\right\},
\end{eqnarray}
where $\Lambda ^{k l}\equiv\epsilon ^{\alpha  \beta } \Lambda ^k{}_{\beta } \Lambda ^l{}_{\alpha }$ that is the product of two BFNC parameters, 
and the various identities given in eq.~(\ref{identities_used}) have been used.

Now we make the transformation for the deformed action from its component form eq.~(\ref{star_of_chiral_field_component_form}) to the desired superfield form. This performance is necessary for us to compute effective actions in section 5.

The component fields can be transformed to the chiral and antichiral superfields as follows,
\begin{eqnarray}
\label{component_to_superfield}
A&=&\Phi |,\qquad \psi _{\alpha }=\frac{1}{\sqrt{2}}D_{\alpha } \Phi |,\qquad F=-\frac{1}{4} \left(D^2 \Phi \right)|,\nonumber\\
A^*&=&\Phi ^+|,\qquad \bar{\psi }_{\dot{\alpha }}=\frac{1}{\sqrt{2}}\bar{D}_{\dot{\alpha }} \Phi ^+|,\qquad F^*=-\frac{1}{4} \left(\bar{D}^2 \Phi ^+\right)|,
\end{eqnarray}
where the symbol $|$ represents setting all $\theta ^{\alpha }$ and $\bar{\theta }^{\dot{\alpha }}$ be zero.

The deformed action ${\cal S}_{\rm NC}$ is the sum of the ordinary part ${\cal S}_{\rm WZ}$ (the terms in the first two lines, also see eq.~(\ref{WZ_superfield_form})) and the noncommutative part ${\cal S}_{{\Lambda}}$ (the remaining terms),
\begin{eqnarray}
\label{S_NC}
{\cal S}_{{\rm NC}}
&=&\int d^8z \left\{\Phi ^+\Phi -\frac{m}{8}\Phi  \left(\frac{D^2}{\square }\Phi \right)-\frac{m}{8}\Phi ^+ \left(\frac{\bar{D}^2}{\square }\Phi ^+\right)\right.\nonumber\\
&&\hspace{15mm}-\frac{g}{12}\Phi  \Phi \left(\frac{D^2}{\square }\Phi \right)-\frac{g}{12}\Phi ^+ \Phi ^+ \left(\frac{\bar{D}^2}{\square }\Phi ^+\right)\nonumber\\
&&\hspace{15mm}+\frac{1}{3072}(-g) \Lambda ^{k l} \Lambda ^{n o} \theta ^4 \left(D^2 \Phi \right) \partial _l\partial _k\left(D^2 \Phi \right) \partial _o\partial _n\left(D^2 \Phi \right)\nonumber\\
&&\hspace{15mm}+\frac{1}{32} (-g) \Lambda ^{k l} \theta ^4 \Phi  \left(D^2 \Phi \right) \partial _l\partial _k\left(D^2 \Phi \right)\nonumber\\
&&\hspace{15mm}+\frac{1}{32} (-g) \Lambda ^{k l} \theta ^4 \Phi  \partial _k\left(D^2 \Phi \right) \partial _l\left(D^2 \Phi \right)\nonumber\\
&&\hspace{15mm}+\frac{1}{6} (-g) \Lambda ^{k l} \theta ^4 \Phi ^+ \square \Phi ^+ \partial _k\partial _l\Phi ^+\nonumber\\
&&\hspace{15mm}+\frac{1}{3} (-g) \left(\sigma \Lambda \Lambda ^{k l}\right)^{n o} \theta ^4 \Phi ^+ \partial _k\partial _n\Phi ^+ \partial _l\partial _o\Phi ^+\nonumber\\
&&\hspace{15mm}+\frac{1}{6} g \eta ^{k l} \Lambda ^{n o} \theta ^4 \Phi ^+ \partial _k\partial _n\Phi ^+ \partial _l\partial _o\Phi ^+\nonumber\\
&&\hspace{15mm}+\frac{1}{16} (-g) \epsilon ^{\alpha  \beta } \Lambda ^{k l} \theta ^4 \partial _k\left(D_{\alpha } \Phi \right) \partial _l\left(D_{\beta } \Phi \right) \left(D^2 \Phi \right)\nonumber\\
&&\hspace{15mm}\left.+\frac{1}{16} (-g) \epsilon ^{\alpha  \beta } \epsilon ^{\zeta  \iota } \Lambda ^k{}_{\beta } \Lambda ^l{}_{\iota } \theta ^4 \partial _k\left(D_{\alpha } \Phi \right) \partial _l\left(D_{\zeta } \Phi \right) \left(D^2 \Phi \right)
\right\},
\end{eqnarray}
where the symbol $\theta ^4$  has been introduced in order to express the above equation concisely, 
\begin{eqnarray}
\theta ^4 &\equiv& \theta ^2\bar{\theta }^2,
\label{newsymbols}
\end{eqnarray}
and the meanings of $\theta ^2$ and $\bar{\theta }^2$ are provided in eq.~(\ref{theta2}).

Note that ${\cal S}_{\rm NC}$ (eq.~(\ref{S_NC})) contains the second and fourth orders of contributions of the BFNC parameters $\Lambda ^{k}{}_{ \alpha }$'s. In addition, it is invariant under the $1/2$ supersymmetry transformation defined by
\begin{equation}
\label{supersymmetry transformation}
\delta _{\xi }\Phi\equiv\xi ^{\alpha }Q_{\alpha }\Phi, \qquad \delta _{\xi }\Phi ^+\equiv\xi ^{\alpha }Q_{\alpha }\Phi ^+.
\end{equation}

At the end of this section, we discuss how to understand nilpotent parameters.
 
The general NC parameters on the canonical NC spacetime~\cite{Seiberg:1999vs}, on the NAC superspace~\cite{Ooguri:2003qp, Ooguri:2003tt, Seiberg:2003yz, Berkovits:2003kj}, and on the BFNC superspace~\cite{de Boer:2003dn} correspond to constant Neveu-Schwarz $B$ fields, graviphoton fields, and gravitino fields, respectively. These fields are solutions of EOM in string theory. The NC parameters on the canonical NC spacetime and the NAC parameters on the NAC superspace are ordinary numbers, while the BFNC parameters on the BFNC superspace are Grassmann numbers that are nilpotent. The general NC parameters will enter into the expressions of physical quantities, such as probabilities of various quantum scattering processes, cross-sections, and so on. How to interpret nilpotent parameters in physical quantities is the key question that we must answer in order to extend models to the BFNC superspace. Grassmann BFNC parameters determine the associativity of star product, and play an important role in the renormalizability of the deformed model~\cite{Miao:2014b}.  If Grassmann parameters are introduced, the only way to cancel them is to do Berezin integration~\cite{Berezin:1966}.

On the canonical NC spacetime (NAC superspace), the NC (NAC) parameters related to $B$ fields (graviphoton fields) are normally regarded as constants. However, as all the constant background fields are solutions of EOM in string theory, $B$ fields (graviphoton fields) can take all allowed constant values. In other words, one can define a probability distribution function for the NC (NAC) parameters related to $B$ fields (graviphoton fields), then take into account the effect of the probability distribution function on physical quantities. If the structure of spacetime (superspace) at a small scale is determined by string theory, the dynamics of string will give the explicit form of the probability distribution function.

The above analysis provides us a clue on how to solve the problem of nilpotent parameters on the BFNC superspace. Following the canonical NC spacetime and NAC superspace, we can introduce the probability distribution function $\rho (\Lambda )$ for BFNC parameters $\Lambda ^k{}_{\alpha }$, which is the generalization of probability distribution function of ordinary numbers.

We denote a general physical quantity related to BFNC parameters $\Lambda ^k{}_{\alpha }$ as $P(\Lambda )$. In order to take into account the effect of the probability distribution function $\rho (\Lambda )$ on $P(\Lambda )$, we define the average value of physical quantity $P(\Lambda )$ by
\begin{equation}
\label{integral_1}
\int d^8\Lambda~\rho (\Lambda )~P(\Lambda ),
\end{equation}
where $d^8\Lambda$ stands for the volume element of 8 components of $\Lambda ^k{}_{\alpha }$, $k=0, 1, 2, 3$, $\alpha=1, 2$. 

We can postulate a general form of $\rho (\Lambda )$ in terms of the power expansion of $\Lambda ^k{}_{\alpha }$,
\begin{equation}
\label{expan_1}
\rho (\Lambda )=A+B_{kl}{}^{\alpha \beta }\Lambda ^k{}_{\alpha }\Lambda ^l{}_{\beta }+{\cdots},
\end{equation}
where ``$\cdots$" represents higher order terms. 
Note that eq.~(\ref{expan_1}) is only needed to be expanded to the eighth order
since there are 8 different nilpotent BFNC parameters $\Lambda ^k{}_{\alpha }$ in total.
Moreover, only even powers of $\Lambda ^k{}_{\alpha }$ in eq.~(\ref{expan_1}) appear, so the coefficients $A$, $B_{{kl}}{}^{\alpha \beta }$, etc. are ordinary numbers. 
The reason is obvious. Because only even powers of the BFNC parameters $\Lambda ^k{}_{\alpha }$ exist in the deformed action, see eq.~(\ref{S_NC}), $P(\Lambda )$ can only contain the terms with even powers of parameters $\Lambda ^k{}_{\alpha }$. As a result, the terms with  odd powers of $\Lambda ^k{}_{\alpha }$ in $\rho (\Lambda )$ have no contributions to the integration in eq.~(\ref{integral_1}). 

After integration in eq.~(\ref{integral_1}), the average value of physical quantity $P(\Lambda )$  will only contain ordinary numbers.

The integration of probability distribution function $\rho (\Lambda )$ corresponds to the total probability and is normalized to be unit, 
\begin{equation}
\label{integral_2}
\int d^8\Lambda~\rho (\Lambda )=1,
\end{equation}
which imposes constraints upon the coefficients of eighth order terms in the expansion eq.~(\ref{expan_1}).

If it is assumed that the physical quantity $P(\Lambda )$ has a proper limit when $\Lambda ^k{}_{\alpha }$ approaches to zero, then $P(\Lambda )$ can be separated into two parts, 
\begin{equation}
\label{physical_quantity}
P(\Lambda )=P_0+P_1(\Lambda),
\end{equation}
where $P_0$ does not contain $\Lambda ^k{}_{\alpha }$ and corresponds to the expression of physical quantities for the undeformed part of our model, and $P_1(\Lambda)$ contains factors of even powers of $\Lambda ^k{}_{\alpha }$ and corresponds to noncommutative corrections. Therefore, eq.~(\ref{integral_1}) can be rewritten as 
\begin{equation}
\label{integral_3}
\int d^8\Lambda~\rho \left(\Lambda \right)\left(P_0+P_1\left(\Lambda\right)\right)=P_0+\int d^8\Lambda~\rho \left(\Lambda \right)P_1\left(\Lambda\right),
\end{equation}
which only contains ordinary numbers as stated above.
As a result, our analysis gives a consistent interpretation of nilpotent parameters.


\section{Background Field Method}
\label{Background_Field_Method}

In this section we give a brief review  on the background field method~\cite{Gates:1983nr,Dimitrijevic:2010yv}.
The purpose is to prepare some necessary formulae, such as the matrices $M$, $M^{-1}$, and $V$, for deriving effective actions related to the deformed Wess-Zumino action ${\cal S}_{\rm NC}$ (eq.~(\ref{S_NC})) on the BFNC superspace in section 5. 

\subsection{General Procedure}

We split the chiral and antichiral superfields into the classical and quantum parts,
\begin{equation}
\label{replace}
\Phi \to \Phi +\Phi _q,\qquad \Phi ^+\to \Phi ^++\Phi _q^+,
\end{equation}
where the quantum parts of the chiral and antichiral superfields satisfy the constraints: $\bar{D}_{\dot{\alpha }}\Phi _q=D_{\alpha }\Phi _q^+=0$.

To integrate out the quantum parts 
we represent them by the general superfields $\Sigma$ and $\Sigma^+$, respectively, 
\begin{equation}
\label{represent quantum part}
\Phi _q=-\frac{1}{4}\bar{D}^2\Sigma,\qquad \Phi _q^+=-\frac{1}{4}D^2\Sigma ^+. 
\end{equation}
Then the quantum fields $\Sigma$ and $\Sigma^+$ can be treated as free fields.

Because of the introduction of the covariant derivative in eq.~(\ref{represent quantum part}) and the nilpotency of operators $D$ and $\bar{D}$, we have the new gauge symmetry,
\begin{equation}
\label{gaugetrans}
\Sigma \to \Sigma +\bar{D}_{\dot{\alpha }}\bar{\Lambda }^{\dot{\alpha }},\qquad \Sigma ^+\to \Sigma ^++D^{\alpha }\Lambda _{\alpha},
\end{equation}
and thus need to add the gauge fixing action ${\cal S}_{{\rm GF}}$,
\begin{equation}
\label{S_GF}
{\cal S}_{\rm GF}=\int d^8z\left\{-\frac{3}{16}\xi  \epsilon ^{\dot{\alpha } \dot{\beta }} \left(\bar{D}_{\dot{\alpha }} \Sigma ^+\right) \left(\bar{D}_{\dot{\beta }} D^2 \Sigma \right)-\frac{1}{4} \xi  \epsilon ^{\alpha  \beta } \epsilon ^{\dot{\zeta } \dot{\iota }} \left(\bar{D}_{\dot{\zeta }} \Sigma ^+\right) \left(D_{\beta } \bar{D}_{\dot{\iota }} D_{\alpha } \Sigma \right) \right\},
\end{equation}
in order to eliminate the degree of freedom in eq.~(\ref{gaugetrans}), where $\xi$ is a gauge fixing parameter.

Replacing the chiral and antichiral superfields in the deformed Wess-Zumino action ${\cal S}_{\rm NC}$ 
(eq.~(\ref{S_NC}))  
by eq.~(\ref{replace}) and keeping only the quadratic part of the quantum fields in the combined action ${\cal S}_{\rm NC}+{\cal S}_{\rm GF}$,  we have
\begin{equation}
\label{matrix}
{\cal S}^{(2)}=\frac{1}{2}\int d^8z\left(
\begin{array}{cc}
 \Sigma  & \Sigma ^+
\end{array}
\right)(M+V)\left(
\begin{array}{c}
 \Sigma  \\
 \Sigma ^+
\end{array}
\right),
\end{equation}
where the matrices $M$ and $V$ correspond to the kinetic and interacting parts of the action ${\cal S}_{\rm NC}+{\cal S}_{\rm GF}$.

After abstracting  the free part in the action ${\cal S}^{(2)}$ and making the Taylor expansion, 
we obtain the one-loop $n$-point effective action $\Gamma^{(n)}$ from the following formula, 
\begin{equation}
\label{EF expand}
\Gamma=\frac{i}{2}{\rm STr} \ln \left(1+M^{-1}V\right)=\frac{i}{2} \sum _{n=1}^{\infty }{\rm STr}\left[ \frac{(-1)^{n+1}}{n}\left(M^{-1}V\right)^n\right]\equiv \sum _{n=1}^{\infty } \Gamma^{(n)}.
\end{equation}

\subsection{Calculation of $M$ and $V$}

Each term in ${\cal S}^{(2)}$ can be represented in a general form,
\begin{equation}
\label{abstract_form}
\int d^8z\left(\partial _1X\right) Z
\left(\partial _2Y\right),
\end{equation}
where $\partial _1$ and $\partial _2$ stand for any products of the following operators,
\begin{equation}
\label{operator_form}
\partial _k, \qquad 
D_{\alpha },\qquad 
\bar{D}_{\dot{\alpha }},\qquad 
\square ^{-1},
\end{equation}
and $X$ and $Y$ denote $\Sigma$  or $\Sigma ^+$,
$Z$ can be the identity operator, $\theta^4$, or a function of chiral and antichiral superfields $\Phi$ and $\Phi^+$.

To obtain $M+V$, we find the following rules for each term in ${\cal S}^{(2)}$,
\begin{itemize}
\label{rule_for_matrix}
\item 
Each $\int d^8z\left(\partial _1\Sigma \right) Z \left(\partial _2\Sigma \right)$ contributes 
$\left(\partial _1{}^TZ\partial _2+\left(\partial _1{}^TZ\partial _2\right){}^T\right)$ to $(M+V)_{1,1}$,

\item 
Each $\int d^8z\left(\partial _1\Sigma \right) Z \left(\partial _2\Sigma ^+\right)$ contributes 
$\partial _1{}^TZ\partial _2$ to $(M+V)_{1,2}$, 
and $\left(\partial _1{}^T Z \partial _2\right){}^T$ to $(M+V)_{2,1}$,

\item 
Each $\int d^8z\left(\partial _1\Sigma ^+\right) Z \left(\partial _2\Sigma ^+\right)$ contributes 
$\left(\partial _1{}^TZ\partial _2+\left(\partial _1{}^TZ\partial _2\right){}^T\right)$ to $(M+V)_{2,2}$,

\end{itemize}
where $(M+V)_{i,j}$ is the $(i, j)$ component of matrix $M+V$.
With these rules we can ensure that the matrices $M$ and $V$ are symmetric, that is, they satisfy $M=M^{T}$ and $V=V^{T}$.

For any operators $A$ and $B$, the transposition of their product is defined by
\begin{equation}
\label{}
(A B)^T\equiv(-1)^{|A\|B|}B^TA^T,
\end{equation}
and the action of $T$ on operators is defined as follows:
\begin{eqnarray}
\label{transpose_rules}
\partial _k{}^T&\equiv&-\partial _k,\qquad \square ^T\equiv\square ,\qquad \left(\frac{1}{\square }\right)^T\equiv\frac{1}{\square },\nonumber\\
\partial _{\alpha }{}^T&\equiv&-\partial _{\alpha },\qquad \bar{\partial }_{\dot{\alpha }}{}^T\equiv-\partial _{\dot{\alpha }},\nonumber\\
D_{\alpha }{}^T&\equiv&-D_{\alpha },\qquad \bar{D}_{\dot{\alpha }}{}^T\equiv-\bar{D}_{\dot{\alpha }},\nonumber\\
\left(D^2\right)^T&\equiv&D^2,\qquad \left(\bar{D}^2\right)^T\equiv\bar{D}^2.
\end{eqnarray}

In accordance with the above considerations, we can easily obtain $M$ and $V$ from $M+V$, where $M$ does not include $\Phi $ and $\Phi ^+$,  but $V$ contains them.
The results are given as follows,
\begin{eqnarray}
\label{Mmatrix}
M_{1,1}&=&-\frac{1}{4} m \bar{D}^2,\nonumber\\
M_{1,2}&=&\frac{1}{16} \bar{D}^2 D^2+\frac{3}{16}\xi  D^2 \bar{D}^2+\frac{1}{4} \xi  \epsilon ^{\alpha  \beta } \epsilon ^{\dot{\zeta } \dot{\iota }} D_{\alpha } \bar{D}_{\dot{\iota }} D_{\beta } \bar{D}_{\dot{\zeta }},\nonumber\\
M_{2,1}&=&\frac{3}{16} \xi  \bar{D}^2 D^2+\frac{1}{16} D^2 \bar{D}^2+\frac{1}{4} \xi  \epsilon ^{\alpha  \beta } \epsilon ^{\dot{\zeta } \dot{\iota }} \bar{D}_{\dot{\zeta }} D_{\beta } \bar{D}_{\dot{\iota }} D_{\alpha },\nonumber\\
M_{2,2}&=&-\frac{1}{4} m D^2,
\end{eqnarray}
and 
{\small
\begin{eqnarray}
\label{matrix_V}
V_{1,2}&=&0,\qquad V_{2,1}=0,\nonumber\\
V_{1,1}&=&\frac{1}{2} (-g) \Phi  \bar{D}^2+\frac{1}{512} (-g) \bar{D}^2 D^2 \theta ^4 \left(D^2 \Phi \right) {p\Lambda }^2 \bar{D}^2+\frac{1}{512} (-g) {p\Lambda }^2 \bar{D}^2 \theta ^4 \left(D^2 \Phi \right) D^2 \bar{D}^2\nonumber\\
&&+\frac{1}{128} (-g) \epsilon ^{\alpha  \beta } \epsilon ^{\zeta  \iota } {p\Lambda }_{\beta } \bar{D}^2 D_{\iota } \theta ^4 \left(D^2 \Phi \right) {p\Lambda }_{\zeta } D_{\alpha } \bar{D}^2+\frac{1}{8192}(-g) \bar{D}^2 D^2 \theta ^4 \left({p\Lambda }^2 \left(D^2 \Phi \right)\right) {p\Lambda }^2 D^2 \bar{D}^2\nonumber\\
&&+\frac{1}{128} (-g) \epsilon ^{\alpha  \beta } \epsilon ^{\zeta  \iota } \bar{D}^2 D^2 \theta ^4 \left({p\Lambda }_{\beta } \left(D_{\iota } \Phi \right)\right) {p\Lambda }_{\zeta } D_{\alpha } \bar{D}^2+\frac{1}{512} (-g) \bar{D}^2 D^2 \theta ^4 \left({p\Lambda }^2 \Phi \right) D^2 \bar{D}^2\nonumber\\
&&+\frac{1}{128} (-g) \epsilon ^{\alpha  \beta } \epsilon ^{\zeta  \iota } {p\Lambda }_{\beta } \bar{D}^2 D_{\iota } \theta ^4 \left({p\Lambda }_{\zeta } \left(D_{\alpha } \Phi \right)\right) D^2 \bar{D}^2,\nonumber\\
V_{2,2}&=&\frac{1}{2} (-g) \Phi ^+ D^2+\frac{1}{96} (-g) D^2 \theta ^4 \left(\square \Phi ^+\right) {p\Lambda }^2 D^2+\frac{1}{96} (-g) {p\Lambda }^2 D^2 \theta ^4 \left(\square \Phi ^+\right) D^2\nonumber\\
&&+\frac{1}{96} (-g) \square  D^2 \theta ^4 \Phi ^+ {p\Lambda }^2 D^2+\frac{1}{96} (-g) {p\Lambda }^2 D^2 \theta ^4 \Phi ^+ \square  D^2+\frac{1}{96} (-g) D^2 \theta ^4 \left({p\Lambda }^2 \Phi ^+\right) \square  D^2\nonumber\\
&&+\frac{1}{48} (-g) \epsilon ^{\alpha  \beta } D^2 \theta ^4 \left({p\sigma }_{\beta  \dot{\zeta }} \left({p\Lambda }_{\iota } \Phi ^+\right)\right) \overline{{p\sigma }}^{\dot{\zeta } \iota } {p\Lambda }_{\alpha } D^2+\frac{1}{48} (-g) \epsilon ^{\alpha  \beta } {p\sigma }_{\beta  \dot{\zeta }} {p\Lambda }_{\iota } D^2 \theta ^4 \Phi ^+ \overline{{p\sigma }}^{\dot{\zeta } \iota } {p\Lambda }_{\alpha } D^2\nonumber\\
&&+\frac{1}{96} (-g) \square  D^2 \theta ^4 \left({p\Lambda }^2 \Phi ^+\right) D^2+\frac{1}{48} (-g) \epsilon ^{\alpha  \beta } {p\sigma }_{\beta  \dot{\zeta }} {p\Lambda }_{\iota } D^2 \theta ^4 \left(\overline{{p\sigma }}^{\dot{\zeta } \iota } \left({p\Lambda }_{\alpha } \Phi ^+\right)\right) D^2,
\end{eqnarray}
}
where the following new symbols have been introduced in order to express matrix $V$ concisely, 
\begin{eqnarray}
{p\sigma}_{\alpha  \dot{\beta }}  &\equiv& \sigma ^k{}_{\alpha  \dot{\beta }} \partial _k,\qquad 
\overline{{p\sigma}}^{\dot{\alpha } \beta } \equiv \left(\bar{\sigma }^k\right)^{\dot{\alpha } \beta } \partial _k ,\nonumber\\
{p\Lambda}^{\alpha } &\equiv& \Lambda ^{k \alpha } \partial _k,\qquad 
{p\Lambda}_{\alpha } \equiv \Lambda ^k{}_{\alpha } \partial _k,\nonumber\\
{p\Lambda }^2&\equiv&{p\Lambda }^{\alpha }{p\Lambda }_{\alpha }.
\label{newsymbols}
\end{eqnarray}

\subsection{Calculation of $M^{-1}$}

We at first briefly review the general procedure to calculate $M^{-1}$~\cite{Wess:1992cp}.
For a general $2 \times 2$ matrix $X$ that can be written as
\begin{equation}
\label{}
X= a P_1+d P_2+b P_++c P_-+e P_T,
\end{equation}
where $a$, $b$, $c$, $d$, and $e$ are $2 \times 2$ coefficient matrices, and $P_1$, $P_2$, $P_+$, $P_-$, and $P_T$ are related to derivatives of the chiral coordinates and defined by
\begin{eqnarray}
\label{}
P_1&\equiv&\frac{1}{16} \square ^{-1} D^2 \bar{D}^2,\qquad P_2\equiv\frac{1}{16} \square ^{-1} \bar{D}^2 D^2,\nonumber\\
P_+&\equiv&\frac{1}{4} \square ^{-\frac{1}{2}} D^2,\qquad P_-\equiv\frac{1}{4} \square ^{-\frac{1}{2}} \bar{D}^2,\nonumber\\
P_T&\equiv&-\frac{1}{8} \epsilon ^{\alpha  \beta } \square ^{-1} D_{\beta } \bar{D}^2 D_{\alpha },
\end{eqnarray}
one  has the inverse of $X$,
\begin{eqnarray}
\label {invers_of_matrix}
X^{-1}&=&\left(a-b d^{-1}c\right)^{-1}P_1+\left(d-c a^{-1}b\right)^{-1}P_2-a^{-1}b\left(d-c a^{-1}b\right)^{-1}P_+\nonumber\\
&&-d^{-1}c\left(a-b d^{-1}c\right)^{-1}P_-+e^{-1}P_T,
\end{eqnarray}
if $a$, $d$, and $e$ are invertible.

Now we turn to our case and  at first transform $M$ (eq.~(\ref{Mmatrix})) to the following form,
\begin{equation}
\label{Mmatrix2}
M=\left(
\begin{array}{cc}
 -m \square ^{\frac{1}{2}} P_- & \xi  \square  P_1+\square  P_2+\xi  \square  P_T \\
 \square  P_1+\xi  \square  P_2+\xi  \square  P_T & -m \square ^{\frac{1}{2}} P_+
\end{array}
\right),
\end{equation}
where we have used the equalities
\begin{eqnarray}
\label{}
\epsilon ^{\alpha  \gamma } \epsilon ^{\dot{\beta } \dot{\zeta }} D_{\alpha } \bar{D}_{\dot{\beta }} D_{\gamma } \bar{D}_{\dot{\zeta }}&=&\frac{1}{2} D^2 \bar{D}^2-\frac{1}{2} \epsilon ^{\alpha  \beta } D_{\alpha } \bar{D}^2 D_{\beta },\nonumber\\
\epsilon ^{\beta  \zeta } \epsilon ^{\dot{\alpha } \dot{\gamma }} \bar{D}_{\dot{\alpha }} D_{\beta } \bar{D}_{\dot{\gamma }} D_{\zeta }&=&\frac{1}{2} \bar{D}^2 D^2-\frac{1}{2} \epsilon ^{\alpha  \beta } D_{\alpha } \bar{D}^2 D_{\beta }.
\end{eqnarray}
Next, by applying eq.~(\ref {invers_of_matrix}) to eq.~(\ref{Mmatrix2}) we thus work out the inverse of $M$,
\begin{eqnarray}
\label{Minverse}
\left(M^{-1}\right){}_{1,1}&=&\frac{1}{4}\frac{m}{\square -m^2} \square ^{-1} D^2,\nonumber\\
\left(M^{-1}\right){}_{1,2}&=&\frac{1}{16 \xi }\square ^{-2} \bar{D}^2 D^2+\frac{1}{16} \frac{1}{\square -m^2} \square ^{-1} D^2 \bar{D}^2-\frac{1}{8 \xi }\epsilon ^{\alpha  \beta } \square ^{-2} D_{\beta } \bar{D}^2 D_{\alpha },\nonumber\\
\left(M^{-1}\right){}_{2,1}&=&\frac{1}{16} \frac{1}{\square -m^2} \square ^{-1} \bar{D}^2 D^2+\frac{1}{16 \xi }\square ^{-2} D^2 \bar{D}^2-\frac{1}{8 \xi }\epsilon ^{\alpha  \beta } \square ^{-2} D_{\beta } \bar{D}^2 D_{\alpha },\nonumber\\
\left(M^{-1}\right){}_{2,2}&=&\frac{1}{4}\frac{m}{\square -m^2} \square ^{-1} \bar{D}^2.
\end{eqnarray}

\section{Main Result}
\label{Main_Result}

\subsection{General Procedure for Calculating Supertrace}

We give the general procedure to compute ${\rm STr}$.
The $n$-point function corresponds to
\begin{equation}
\label{n_point}
\frac{i(-1)^{n+1}}{2 n}{\rm STr}\left(M^{-1}V\right)^n,
\end{equation}
where ${\rm STr}$ is the supertrace. The whole process can be divided into two parts, the first is to calculate the trace for the spinor coordinate $\theta^4=\theta ^2\bar{\theta }^2$, and the second part is to calculate the trace for the Bosonic coordinate $x^k$.

Besides the BFNC parameter factors, we can see from eqs.~(\ref{Minverse}) and (\ref{matrix_V}) that the general form of the terms in $M^{-1}V$ is
\begin{equation}
\label{general_form_MIV}
{\cal D}_A
\partial _C
\frac{1}{\square -m^2}
\left(\square ^{-1}\right)^{s}
\theta^4
{\cal F} _B
\partial _{C'}
{\cal D}_{A'},
\end{equation}
where $s=0$ in some terms which means $\square ^{-1}$ does not appear, or $s=1$ in some terms which means $\square ^{-1}$ appears. 
We explain the meanings of the symbols in eq.~(\ref{general_form_MIV}) as follows.

${\cal D}_A$ and ${\cal D}_{A'}$ stand for any elements of set ${\cal D}$,
\begin{equation}
\label{set_D}
{\cal D}=\left\{D_{\alpha },\,
\bar{D}_{\dot{\beta }},\,
D^2,\,
\bar{D}^2,\,
D_{\alpha }\bar{D}_{\dot{\beta }},\,
D^2\bar{D}_{\dot{\beta }},\,
D_{\alpha }\bar{D}^2,\,
D^2\bar{D}^2\right\}.
\end{equation}
To obtain the set ${\cal D}$, we have taken into account the $D$ algebraic relations eq.~(\ref{D_algebraic_relation}) and move every $D_{\alpha }$ to the left side of $\bar{D}_{\dot{\beta }}$.

$\partial _C$ and $\partial _{C'}$ denote any elements of set ${\cal \partial}$,
\begin{equation}
\label{}
{\cal \partial}=\left\{
\partial _k,\,
\partial _k\partial _l,\,
\partial _k\partial _l\partial _m, \,\cdots\right\}.
\end{equation}

${\cal F} _B$ is any element of set ${\cal F}$,
\begin{equation}
\label{set_F}
{\cal F}=\left\{
\left((\partial )^n\Phi \right), \,\left((\partial )^nD_{\alpha }\Phi \right), \,\left((\partial )^nD^2\Phi \right), \,\left((\partial )^n\Phi ^+\right),\,  \left((\partial )^n\bar{D}_{\dot{\alpha }}\Phi ^+\right)\left((\partial )^n\bar{D}^2\Phi ^+\right)
\right\},
\end{equation}
where $(\partial )^n$ implies the product of $n$ $\partial _k$'s,
\begin{equation}
\label{}
(\partial )^n=\partial _{k_1}\partial _{k_2}\cdots\partial _{k_n},
\end{equation}
and $n$ is a non-negative integer.
To obtain ${\cal F}$, we have used the $D$ algebraic relations eq.~(\ref{D_algebraic_relation}) and the chiral and antichiral conditions eq.~(\ref{chiralcondition}).

From eq.~(\ref{general_form_MIV}), we can determine the general form of eq.~(\ref{n_point}),
\begin{equation}
\label{general_form_STr}
{\rm STr}\left\{\underbrace{\left(
{\cal D}_{A_1}\partial _{C_1}\frac{1}{\square -m^2}\left(\square ^{-1}\right)^{s_1}\theta^4{\cal F}_{B_1}\partial _{C^{\prime}_1}{\cal D}_{A^{\prime}_1}\right)\cdots
\left({\cal D}_{A_n}\partial _{C_n}\frac{1}{\square -m^2}\left(\square ^{-1}\right)^{s_n}\theta^4{\cal F}_{B_n}\partial _{C^{\prime}_n}{\cal D}_{A^{\prime}_n}\right)}_{\rm{n\,terms}}\right\},
\end{equation}
where ${\cal D}_{A_i}$ and ${\cal D}_{A^{\prime}_i}$ stand for any elements of set ${\cal D}$, ${\cal F}_{B_i}$ means any element of set $\cal F$,  ${\partial}_{C_i}$ and ${\partial}_{C^{\prime}_i}$ denote any elements of set ${\cal \partial}$, 
and $s_i$ takes the value of $0$ or $1$, $i=1, 2, \cdots, n$.

We use the following procedure to show how to calculate eq.~(\ref{general_form_STr}), for simplicity, but without the loss of generality, we take $n=2$ as a sample.

{\em Step 1}:  By using the symmetry of supertrace ${\rm STr}$,
\begin{equation}
\label{}
{\rm STr}\{{\cal X} {\cal Y}\}=(-1)^{|{\cal X}\|{\cal Y}|}{\rm STr}\{{\cal Y} {\cal X}\},
\end{equation}
where ${\cal X}$ and ${\cal Y}$ are any operators,
we move $\partial _{C^{\prime}_2}{\cal D}_{A^{\prime}_2}$ to the front of ${\cal D}_{A_1}\partial _{C_1}$ in eq.~(\ref{general_form_STr}) with our choice of $n=2$, and then transform this supertrace to the following form,
\begin{equation}
\label{general_form_STr_rotate_theta}
(-1)^{|{\cal X}\|{\cal Y}|}{\rm STr}\left\{
\partial _{C^{\prime}_2}{\cal D}_{A^{\prime}_2}
\left({\cal D}_{A_1}\partial _{C_1}\frac{1}{\square -m^2}\left(\square ^{-1}\right)^{s_1}\theta^4{\cal F}_{B_1}\partial _{C^{\prime}_1}{\cal D}_{A^{\prime}_1}\right)
{\cal D}_{A_2}\partial _{C_2}\frac{1}{\square -m^2}\left(\square ^{-1}\right)^{s_2}\theta^4{\cal F}_{B_2}\right\},
\end{equation}
where ${\cal Y}=\partial_{C^{\prime}_2}{\cal D}_{A^{\prime}_2}$ and ${\cal X}$ represents the whole factor after $\partial_{C^{\prime}_2}{\cal D}_{A^{\prime}_2}$ in eq.~(\ref{general_form_STr_rotate_theta}).

{\em Step 2}:  We define the Leibniz rule as
\begin{equation}
\label{Leibniz_rule}
{\cal O} {\cal G}=
\left({{\cal O}{\cal G}}\right)
+(-1)^{\left|{\cal O}\left\|{\cal G}\right.\right|}{\cal G}{\cal O},
\end{equation}
where ${\cal O}$ is an element of set $\left\{\partial _k, \,D_{\alpha }, \,\bar{D}_{\dot{\alpha }}\right\}$,
${\cal G}$ denotes an  element of set ${\cal F}$, or $\theta^4$, or $\left({\cal D}_{A}\theta^4\right)$, and
$\left({{\cal O}{\cal G}}\right)$ stands for  ${\cal O}$ acting on ${\cal G}$.

By using the Leibniz rule (eq.~(\ref{Leibniz_rule})) and considering the fact that $D$ commutes with $\partial_k$, we move all $D$ operators to the front of $\theta^4{\cal F}_{B_2}$.
At the same time, by using the $D$ algebraic relations eq.~(\ref{D_algebraic_relation}), we can simplify the covariant derivatives to the forms as the elements in set ${\cal D}$ (eq.~(\ref{set_D})).
Thus, eq.~(\ref{general_form_STr_rotate_theta}) changes to be 
\begin{equation}
\label{general_form_STr_move_D}
{\rm STr}\left\{
\partial _{C^{\prime}_2}
\partial _{C_1}\frac{1}{\square -m^2}\left(\square ^{-1}\right)^{s_1}
\left({\cal D}_{A^{\prime\prime}_1}\theta^4\right)
{\cal F}_{B^{\prime}_1}\partial _{C^{\prime}_1}
\partial _{C_2}\frac{1}{\square -m^2}\left(\square ^{-1}\right)^{s_2}{\cal D}_{A^{\prime\prime}_2}\theta^4{\cal F}_{B_2}\right\},
\end{equation}
where ${\cal F}_{B^{\prime}_1}$ is also an element of set ${\cal F}$,  and ${\cal D}_{A^{\prime\prime}_1}$ and ${\cal D}_{A^{\prime\prime}_2}$ are also elements of set ${\cal D}$. Note that  ${\cal F}_{B^{\prime}_1}$, ${\cal D}_{A^{\prime\prime}_1}$, and ${\cal D}_{A^{\prime\prime}_2}$ can be determined in terms of the $D$ algebraic relations eq.~(\ref{D_algebraic_relation}) and the factor $\left({\cal D}_{A^{\prime\prime}_1}\theta^4\right)$ appears in eq.~(\ref{general_form_STr_move_D}).

{\em Step 3}:  The non-vanishing contribution of eq.~(\ref{general_form_STr_move_D}) requires ${\cal D}_{A^{\prime\prime}_1}=D^2\bar{D}^2$. 
Therefore, eq.~(\ref{general_form_STr_move_D}) reads
\begin{equation}
\label{general_form_STr_move_D_evaluate_theta}
{\rm STr}\left\{
\partial _{C^{\prime}_2}
\partial _{C_1}\frac{1}{\square -m^2}\left(\square ^{-1}\right)^{s_1}
{\cal F}_{B^{\prime}_1}\partial _{C^{\prime}_1}
\partial _{C_2}\frac{1}{\square -m^2}\left(\square ^{-1}\right)^{s_2}{\cal D}_{A^{\prime\prime}_2}\theta^4{\cal F}_{B_2}\right\}.
\end{equation}

{\em Step 4}:   When we calculate  the supertrace of the Fermionic coordinates of eq.~(\ref{general_form_STr_move_D_evaluate_theta}),
its non-vanishing contribution  requires ${\cal D}_{A^{\prime\prime}_2}=D^2\bar{D}^2$.
Thus, eq.~(\ref{general_form_STr_move_D_evaluate_theta}) reduces to
\begin{equation}
\label{general_form_STr_cancel_D}
{\rm Tr}\left\{
\partial _{C^{\prime}_2}
\partial _{C_1}\frac{1}{\square -m^2}\left(\square ^{-1}\right)^{s_1}
{\cal F}_{B^{\prime}_1}\partial _{C^{\prime}_1}
\partial _{C_2}\frac{1}{\square -m^2}\left(\square ^{-1}\right)^{s_2}\theta^4{\cal F}_{B_2}\right\},
\end{equation}
where no covariant derivatives $D$ exist in eq.~(\ref{general_form_STr_cancel_D}).

We make a comment to eq.~(\ref{general_form_STr_cancel_D}) that there must be one $\theta^4$ left in each term containing noncommutative parameters. The following explanation is also valid to the case of a higher (than second) order of noncommutative parameters. Because each term of the deformed part of the Wess-Zumino action contains one $\theta^4$, there is at least one $\theta^4$ in each term associated with noncommutative parameters after expanding ${\rm STr}[(M^{-1}V)^n]$. For such a term, by using the symmetry of ${\rm STr}$, we move one $\theta^4$ and the superfield behind it to the rightmost end of the term. By using Leibniz rules, we move all of the covariant derivatives $D$'s and $\bar D$'s in front of the above mentioned $\theta^4$. We also use $D$ algebraic relations to simplify the products of covariant derivatives. 
When we evaluate the supertrace ${\rm STr}$, only terms with two $D$'s and two $\bar D$'s are non-vanishing. 
All of the $\theta^4$'s except the rightmost one are deleted by $D^2\bar D^2$'s, at the end there is only one $\theta^4$ left in each term containing noncommutative parameters.

{\em Step 5}: Finally, we compute the momentum integral in eq.~(\ref{general_form_STr_cancel_D}) by using the dimensional regularization method, and adopt the minimal subtraction scheme where only divergent parts remain.

The effective actions we obtain are given in Appendix B with our specific method of classification called the $1/2$ supersymmetry invariant subsets and bases. For the details of the subsets and bases, see subsections 5.8, 5.9, and 5.10.

\subsection{Checking of $1/2$ Supersymmetry Invariance}

We provide in this subsection  the method to check the $1/2$ supersymmetry invariance of effective actions.

Besides the BFNC parameter factors, the general form of effective actions reads
\begin{equation}
\label{action_general_form}
{\cal A}=\int d^8z\, \theta ^4\,\{{\cal F}_A \,{\cal F}_B\,\cdots\},
\end{equation}
where ${\cal F}_A$ and ${\cal F}_B$ are superfields and elements of set ${\cal F}$ (see eq.~(\ref{set_F})), and $\{{\cal F}_A \,{\cal F}_B\,\cdots\}$ represents a product of any number of elements in ${\cal F}$. 
The supersymmetry transformation obeys the Leibniz rule. We now investigate the supersymmetry transformation of an effective action in which there are two superfields,
\begin{equation}
\label{variantion_of_A}
\delta _{\xi }{\cal A}=\int d^8z \,\theta ^4\left\{\left(\delta _{\xi }{\cal F}_A\right) {\cal F}_B+{\cal F}_A\left(\delta _{\xi }{\cal F}_B\right)\right\}.
\end{equation}

For example, if we choose ${\cal F}_A=\left((\partial )^nD_{\alpha }\Phi \right)$, we have 
$\delta _{\xi }{\cal F}_A=\left((\partial )^nD_{\alpha }\delta _{\xi }\Phi \right)$.
Considering the transformation of superfield $\Phi$, $\delta _{\xi }\Phi =\xi ^{\beta }Q_{\beta }\Phi$, 
where $\xi ^{\beta }$ is a constant with spinor indices,
we then obtain $\delta _{\xi }{\cal F}_A=\left((\partial )^nD_{\alpha }\xi ^{\beta }Q_{\beta }\Phi \right)$.
By using the algebraic relation, $\left\{D_{\alpha },Q_{\beta }\right\}=0$, 
and taking into account the fact that $\xi ^{\beta }$ is a constant, we move $Q_{\beta }$ to the  front of $D_{\alpha }$ and write $\delta _{\xi }{\cal F}_A$ as follows,
\begin{equation}
\label{}
\delta _{\xi }{\cal F}_A=\left(\xi ^{\beta }(\partial )^nQ_{\beta }D_{\alpha }\Phi \right).
\end{equation}
Using the identity that $Q_{\beta }$ is equal to $D_{\beta }$ under the superspace integral $\int d^8z \,\theta ^4$,
\begin{equation}
\label{}
\int d^8z \,\theta ^4\left\{\left(\xi ^{\beta }(\partial )^nQ_{\beta }D_{\alpha }\Phi \right){\cal F}_B\right\}=\int d^8z \,\theta ^4\left\{\left(\xi ^{\beta }(\partial )^nD_{\beta }D_{\alpha }\Phi \right){\cal F}_B\right\},
\end{equation}
we thus transform covariant derivatives to the simplest form by using the $D$ algebraic relations.  At the end we transform $\delta _{\xi }{\cal A}$ to the general form of effective actions (see eq.~(\ref{action_general_form})).
If the contributions from the two terms in eq.~(\ref{variantion_of_A}) cancel each other, ${\cal A}$ is invariant under the $1/2$ supersymmetry transformation. In this way, we have checked the effective actions (see our very long final result in Appendix B) and confirmed that our effective actions possess the $1/2$ supersymmetry invariance.

\subsection{New Notations for Presenting Effective Actions}

We define the new symbols for presenting effective actions in a concise form,  
\begin{eqnarray}
\label{}
\Lambda ^{k l}&\equiv&\epsilon ^{\alpha  \beta }\Lambda ^k{}_{\beta }\Lambda ^l{}_{\alpha },\nonumber\\
\Lambda ^2&\equiv&\eta _{k l}\Lambda ^{k l},\nonumber\\
\sigma \Lambda \Lambda &\equiv&\eta _{k n}\eta _{l o}\left(\sigma ^{k l}\right)^{\alpha  \beta } \Lambda ^n{}_{\alpha } \Lambda ^o{}_{\beta },\nonumber\\
\left(\sigma \Lambda ^{k l}\right)^{n \alpha }&\equiv&\left(\sigma ^{k l}\right)^{\beta  \alpha } \Lambda ^n{}_{\beta },\nonumber\\
\left(\eta \sigma \Lambda ^k\right)^{\alpha }&\equiv&\eta _{l n} \left(\sigma ^{n k}\right)^{\beta  \alpha } \Lambda ^l{}_{\beta },\nonumber\\
\left(\eta \sigma \Lambda \Lambda ^k\right)^l&\equiv&\eta _{n o} \left(\sigma ^{o k}\right)^{\alpha  \beta } \Lambda ^n{}_{\alpha } \Lambda ^l{}_{\beta },\nonumber\\
\left(\sigma \Lambda \Lambda ^{k l}\right)^{n o}&\equiv&\left(\sigma ^{k l}\right)^{\alpha  \beta } \Lambda ^n{}_{\alpha } \Lambda ^o{}_{\beta }.
\end{eqnarray}

In addition, we hide the superscripts and/or subscripts of Bosonic derivatives, but only show the number of their product, for example,  $\partial_k \partial_l$ is written as $\partial \partial$, and also hide the subscripts of Fermionic derivatives, such as $D$ and $\bar{D}$  denoting $D_{\alpha}$ and $\bar{D}_{\dot{\beta}}$, respectively. In particular, for all terms in effective actions
we pick out different BFNC parameter factors as one class and different operator factors as the other class, which gives the reader an explicit outline of effective actions. That is, any term in effective actions consists of the product of one class of BFNC parameter factors and one class of operator factors, and all the possible combinations of different BFNC parameter factors and different operator factors give an effective action. For the details,  the reader can confer Appendix B.

As an example, considering the above notations we present ${\cal S}_{\rm NC}$ (eq.~(\ref{S_NC})) by separating it into order of $\Lambda^2$ and order of $\Lambda^4$  as follows.
Note that $\int d^8z \,\theta ^4$ is omitted. 

\subsubsection{Order of $\Lambda^2$}
\begin{itemize}
\item  There exist five different BFNC parameter factors,
\begin{equation}
\label{}
\Lambda ^{k l},\qquad \left(\sigma \Lambda \Lambda ^{k l}\right)^{n o},\qquad \eta ^{k l} \Lambda ^{n o},\qquad \epsilon ^{\alpha  \beta } \Lambda ^{k l},\qquad \epsilon ^{\alpha  \beta } \epsilon ^{\zeta  \iota } \Lambda ^k{}_{\beta } \Lambda ^l{}_{\iota }.
\end{equation}

\item   There exist three different operator factors,
\begin{equation}
\label{}
\partial \partial \Phi  \left(D^2 \Phi \right) \left(D^2 \Phi \right),\qquad
\partial \partial (D \Phi ) (D \Phi ) \left(D^2 \Phi \right),\qquad
\partial \partial \partial \partial \Phi ^+ \Phi ^+ \Phi ^+,
\end{equation}
where the product of several $\partial$'s in the front means it has actions to all the superfields behind it. 

\end{itemize}

\subsubsection{Order of $\Lambda^4$}
\begin{itemize}
\item There exists only one BFNC parameter factor,
\begin{equation}
\label{}
\Lambda ^{k l} \Lambda ^{n o}.
\end{equation}

\item There exists only one operator factor,
\begin{equation}
\label{}
\partial \partial \partial \partial \left(D^2 \Phi \right) \left(D^2 \Phi \right) \left(D^2 \Phi \right).
\end{equation}

\end{itemize}

\subsection{Formulation and Classification of $\Gamma _{\rm 1st}$}

As the first step of searching the renormalizable Wess-Zumino model on the BFNC superspace, we calculate the effective action of ${\cal S}_{\rm NC}$ by using the background field method introduced in section 4, and denote it as $\Gamma _{\rm 1st}$.
Incidentally, the order of $\Lambda ^0$ corresponds to the effective action of the Wess-Zumino model on the ordinary superspace and it does not need to be considered. 
In the following we give $\Gamma _{\rm 1st}$ in terms of the new notations and the method of classification proposed in  subsection 5.3 in order to present the primary effective action in a concise form.
We note that in some order of $\Lambda$ a  certain point function does not appear, which means it has no contribution to $\Gamma _{\rm 1st}$. For instance, at the order of $\Lambda^2$ (see below) the 5- and 6-point functions do not appear, which implies that  they have no contributions to $\Gamma _{\rm 1st}$.
As mentioned in subsection 5.2, we verify that $\Gamma _{\rm 1st}$ is invariant under the $1/2$ supersymmetry transformation, and
we see explicitly  that  $\Gamma _{\rm 1st}$ cannot be absorbed by ${\cal S}_{\rm NC}$ (eq.~(\ref{S_NC})) because $\Gamma _{\rm 1st}$ contains many extra terms that do not exist in ${\cal S}_{\rm NC}$.
Note that we remove the divergent coefficient $\frac {1} {\epsilon}$ from $\Gamma _{\rm 1st}$ and still use $\Gamma _{\rm 1st}$ to denote the primary effective action. 

\subsubsection{Order of $\Lambda^2$}
\begin{itemize}
\item There exist fourteen different BFNC parameter factors,
\begin{eqnarray}
\label{}
&&\left(\eta \sigma \Lambda \Lambda ^k\right)^l,\qquad \Lambda ^2,\qquad \Lambda ^{k l},\qquad \epsilon ^{\alpha  \beta } \left(\eta \sigma \Lambda \Lambda ^k\right)^l,\qquad \epsilon ^{\alpha  \beta } \Lambda ^{k l},\nonumber\\
&&\Lambda ^2 \eta ^{k l},\qquad \Lambda ^2 \epsilon ^{\alpha  \beta },\qquad \Lambda ^2 \epsilon ^{\dot{\alpha } \dot{\beta }},\qquad \Lambda ^2 \left(\bar{\sigma }^k\right)^{\dot{\alpha } \beta },\qquad \Lambda ^2 \eta ^{k l} \epsilon ^{\alpha  \beta },\nonumber\\
&&\Lambda ^{k l} \eta _{l n} \left(\bar{\sigma }^n\right)^{\dot{\alpha } \beta },\qquad \eta _{k l} \left(\bar{\sigma }^l\right)^{\dot{\alpha } \beta } \left(\eta \sigma \Lambda \Lambda ^k\right)^n,\qquad \epsilon ^{\alpha  \beta } \epsilon ^{\zeta  \iota } \Lambda ^k{}_{\beta } \Lambda ^l{}_{\iota },\qquad \epsilon ^{\alpha  \beta } \eta _{k l} \left(\sigma \Lambda ^{l n}\right)^{o \zeta } \Lambda ^k{}_{\beta }.
\end{eqnarray}

\item There exist three different operator factors,
\begin{enumerate}

\item The 2-point function contains four different forms,
\begin{equation}
\label{}
\partial \partial \Phi  \left(D^2 \Phi \right),\qquad
\partial \partial (D \Phi ) (D \Phi ),\qquad
\partial \partial \left(D^2 \Phi \right) \Phi ^+,\qquad
\partial \partial \partial \partial \left(D^2 \Phi \right) \Phi ^+;
\end{equation}

\item The 3-point function contains five different forms,
\begin{eqnarray}
\label{}
&&\left(D^2 \Phi \right) \left(D^2 \Phi \right) \left(\bar{D}^2 \Phi ^+\right),\qquad
\partial (D \Phi ) \left(D^2 \Phi \right) \left(\bar{D} \Phi ^+\right),\qquad
\partial \partial \Phi  \left(D^2 \Phi \right) \Phi ^+,\nonumber\\
&&\partial \partial (D \Phi ) (D \Phi ) \Phi ^+,\qquad
\partial \partial \left(D^2 \Phi \right) \Phi ^+ \Phi ^+;
\end{eqnarray}

\item The 4-point function contains five different forms,
\begin{eqnarray}
\label{}
&&\left(D^2 \Phi \right) \left(D^2 \Phi \right) \left(\bar{D} \Phi ^+\right) \left(\bar{D} \Phi ^+\right),\qquad
\left(D^2 \Phi \right) \left(D^2 \Phi \right) \Phi ^+ \left(\bar{D}^2 \Phi ^+\right),\nonumber\\
&&\partial (D \Phi ) \left(D^2 \Phi \right) \Phi ^+ \left(\bar{D} \Phi ^+\right),\qquad
\partial \partial \Phi  \left(D^2 \Phi \right) \Phi ^+ \Phi ^+,\qquad
\partial \partial (D \Phi ) (D \Phi ) \Phi ^+ \Phi ^+.
\end{eqnarray}

\end{enumerate}
\end{itemize}

\subsubsection{Order of $\Lambda^4$}
\begin{itemize}
\item There exist seven different BFNC parameter factors,
\begin{eqnarray}
\label{}
&&\Lambda ^2 \Lambda ^{k l},\qquad \Lambda ^2 \left(\sigma \Lambda \Lambda ^{k l}\right)^{n o},\qquad \Lambda ^{k l} \left(\eta \sigma \Lambda \Lambda ^n\right)^o,\qquad \Lambda ^{k l} \Lambda ^{n o},\qquad \Lambda ^2 \eta ^{k l} \Lambda ^{n o},\nonumber\\
&&\Lambda ^2 \epsilon ^{\alpha  \beta } \Lambda ^{k l},\qquad \Lambda ^{k l} \eta _{l n} \left(\sigma \Lambda \Lambda ^{n o}\right)^{p q}.
\end{eqnarray}

\item There exist five different operator factors,
\begin{enumerate}

\item The 2-point function contains two different forms,
\begin{equation}
\label{}
\partial \partial \left(D^2 \Phi \right) \left(D^2 \Phi \right),\qquad 
\partial \partial \partial \partial \left(D^2 \Phi \right) \left(D^2 \Phi \right);
\end{equation}

\item The 3-point function contains four different forms,
\begin{eqnarray}
\label{}
&&\partial \partial \Phi  \left(D^2 \Phi \right) \left(D^2 \Phi \right),\qquad 
\partial \partial (D \Phi ) (D \Phi ) \left(D^2 \Phi \right),\qquad 
\partial \partial \left(D^2 \Phi \right) \left(D^2 \Phi \right) \Phi ^+,\nonumber\\
&&\partial \partial \partial \partial \left(D^2 \Phi \right) \left(D^2 \Phi \right) \Phi ^+;
\end{eqnarray}

\item The 4-point function contains four different forms,
\begin{eqnarray}
\label{}
&&\partial \partial \Phi  \left(D^2 \Phi \right) \left(D^2 \Phi \right) \Phi ^+,\qquad 
\partial \partial (D \Phi ) (D \Phi ) \left(D^2 \Phi \right) \Phi ^+,\qquad 
\partial \partial \left(D^2 \Phi \right) \left(D^2 \Phi \right) \Phi ^+ \Phi ^+,\nonumber\\
&&\partial \partial \partial \partial \left(D^2 \Phi \right) \left(D^2 \Phi \right) \Phi ^+ \Phi ^+;
\end{eqnarray}

\item The 5-point function contains three different forms,
\begin{eqnarray}
\label{}
&&\partial \partial \Phi  \left(D^2 \Phi \right) \left(D^2 \Phi \right) \Phi ^+ \Phi ^+,\qquad 
\partial \partial (D \Phi ) (D \Phi ) \left(D^2 \Phi \right) \Phi ^+ \Phi ^+,\nonumber \\ 
&&\partial \partial \left(D^2 \Phi \right) \left(D^2 \Phi \right) \Phi ^+ \Phi ^+ \Phi ^+;
\end{eqnarray}

\item The 6-point function contains two different forms,
\begin{equation}
\label{}
\partial \partial \Phi  \left(D^2 \Phi \right) \left(D^2 \Phi \right) \Phi ^+ \Phi ^+ \Phi ^+,\qquad 
\partial \partial (D \Phi ) (D \Phi ) \left(D^2 \Phi \right) \Phi ^+ \Phi ^+ \Phi ^+.
\end{equation}

\end{enumerate}
\end{itemize}

\subsubsection{Order of $\Lambda^6$}
\begin{itemize}
\item There exists only one BFNC parameter factor,
\begin{equation}
\label{}
\Lambda ^2 \Lambda ^{k l} \Lambda ^{n o}.
\end{equation}

\item There exist three different operator factors,

\begin{enumerate}

\item The 4-point function contains only one form,
\begin{equation}
\label{}
\partial \partial \partial \partial \left(D^2 \Phi \right) \left(D^2 \Phi \right) \left(D^2 \Phi \right) \Phi ^+;
\end{equation}

\item The 5-point function contains only one form,
\begin{equation}
\label{}
\partial \partial \partial \partial \left(D^2 \Phi \right) \left(D^2 \Phi \right) \left(D^2 \Phi \right) \Phi ^+ \Phi ^+;
\end{equation}

\item The 6-point function contains only one form,
\begin{equation}
\label{}
\partial \partial \partial \partial \left(D^2 \Phi \right) \left(D^2 \Phi \right) \left(D^2 \Phi \right) \Phi ^+ \Phi ^+ \Phi ^+.
\end{equation}

\end{enumerate}
\end{itemize}

\subsection{Restriction to Order of $\Lambda ^2$}

Before continuing our analysis, let us temporally recall the corresponding situation for the NAC case~\cite{Grisaru:2003fd} where the effective action of the Wess-Zumino model contains only one term,
\begin{eqnarray}
\label{}
\int d^8z~ U\left(D^2\Phi \right)^2,\qquad U=\theta ^2\bar{\theta }^2C^2,
\end{eqnarray}
where $C$ is the NAC parameter.
In that case the renormalizable action can be obtained by adding the effective action to the deformed Wess-Zumino model on the NAC superspace, where Feynman graphs are used to show how the divergences are produced by the vertices of the action.

We turn to our case on the BFNC superspace, where the primary effective action $\Gamma _{\rm 1st}$ cannot be absorbed by the deformed action (eq.~(\ref{S_NC})). 
In our case the number of correction terms that  should be added to ${\cal S}_{\rm NC}$ (eq.~(\ref{S_NC})) is very large, for instance, $\Gamma _{\rm 1st}$ contains 68 terms only at the order of $\Lambda ^2$.
So, it is a tremendously exciting challenge to find out the successive effective actions needed.

In general, if a model can be modified to be renormalizable, it must be renormalizable at each order of $\Lambda$. On the BFNC superspace the possible orders of $\Lambda$ are even, that is,  $\Lambda ^2$, $\Lambda ^4$, $\Lambda ^6$, and $\Lambda ^8$, and higher orders than $\Lambda ^8$ must be vanishing due to the Fermionic number of the BFNC parameters, where the lowest order of approximation is $\Lambda ^2$.
In order to obtain the basic information of the renormalization property of the Wess-Zumino model on the BFNC superspace, we restrict our analysis only at the order of $\Lambda ^2$ in the following calculations of the successive effective actions, such as $\Gamma _{\rm 2nd}$, $\Gamma _{\rm 3rd}$, and $\Gamma _{\rm 4th}$.

\subsection{Formulation and Classification of $\Gamma _{\rm 2nd}$}

\subsubsection{Method}

We start from ${\cal S}_{\rm NC} + \Gamma _{\rm 1st}$ and calculate its effective action called the secondary effective action $\Gamma _{\rm 2nd}$ up to the order of $\Lambda ^2$. The detailed procedure is as follows.

In eq.~(\ref{S_NC}), ${\cal S}_{\rm NC}$ is written as the sum of the ordinary part ${\cal S}_{\rm WZ}$ and the noncommutative part ${\cal S}_{\Lambda }$, ${\cal S}_{\rm NC}={\cal S}_{\rm WZ}+{\cal S}_{\Lambda }$.
Here we define
\begin{equation}
\label{action_S_1}
{\cal S}_{(1)}\equiv{\cal S}_{\rm WZ}+{\cal S}_{\Lambda }\left(\Lambda ^2\right)+\Gamma _{\rm 1st}\left(\Lambda ^2\right),
\end{equation}
where ${\cal S}_{\Lambda }\left(\Lambda ^2\right)$ stands for the $\Lambda ^2$ part of ${\cal S}_{\Lambda }$, and $\Gamma _{\rm 1st}\left(\Lambda ^2\right)$ for the $\Lambda ^2$ part of   $\Gamma _{\rm 1st}$. ${\cal S}_{\Lambda }\left(\Lambda ^2\right)$ is given in subsection 5.3.1 in a concise form or in eq.~(\ref{S_NC}) in detail, and $\Gamma _{\rm 1st}\left(\Lambda ^2\right)$ is also known, see subsection 5.4.1 in a concise form or Appendix B in detail. Our goal is $\Gamma _{\rm 2nd}\left(\Lambda ^2\right)$ which is the effective action of ${\cal S}_{(1)}$.

In fact, the effective action contributed by the vertices of ${\cal S}_{\Lambda }\left(\Lambda ^2\right)$ is just $\Gamma _{\rm 1st}\left(\Lambda ^2\right)$ which is already known, so  we only need to compute the effective action of ${\cal S}_{\rm WZ}+\Gamma _{\rm 1st}\left(\Lambda ^2\right)$.

By following the general procedure of the background field method, 
we divide ${\cal S}_{\rm WZ}$ (eq.~(\ref{WZ_superfield_form})) into the free and interacting parts, ${\cal S}_{\rm WZ}={\cal S}_0+{\cal S}_{{\rm int}}$,
and still use ${\cal S}_{{\rm GF}}$ (see eq.~(\ref{S_GF})) as the gauge fixing term.

As in calculating $\Gamma _{\rm 1st}$, the matrix $M$ corresponding to $\Gamma _{\rm 2nd}$ is still determined by ${\cal S}_0+{\cal S}_{{\rm GF}}$.
However, different from the case of $\Gamma _{\rm 1st}$, a new matrix $V_{(1)}$ related to the interacting part for $\Gamma _{\rm 2nd}$ is determined by ${\cal S}_{{\rm int}}+\Gamma _{\rm 1st}\left(\Lambda ^2\right)$. In addition, we put  the 2-point function of $\Gamma _{\rm 1st}\left(\Lambda ^2\right)$ into $V_{(1)}$ rather than into $M$ in order to avoid $M$ being not invertible. With all these considerations, we can obtain $\Gamma _{\rm 2nd}\left(\Lambda ^2\right)$.

\subsubsection{Result}

By comparing the BFNC parameter factors and operator ones of $\Gamma _{\rm 2nd}\left(\Lambda ^2\right)$ with that of ${\cal S}_{(1)}$, we find some new forms that do not appear in ${\cal S}_{(1)}$.
We list them as follows.

\begin{itemize}
\item There exist thirteen different BFNC parameter factors,
\begin{eqnarray}
\label{}
&&\sigma \Lambda \Lambda ,\qquad \Lambda ^2 \left(\sigma ^{k l}\right)^{\alpha  \beta },\qquad \sigma \Lambda \Lambda  \eta ^{k l},\qquad \sigma \Lambda \Lambda  \epsilon ^{\alpha  \beta },\qquad \sigma \Lambda \Lambda  \epsilon ^{\dot{\alpha } \dot{\beta }},\qquad\sigma \Lambda \Lambda  \left(\bar{\sigma }^k\right)^{\dot{\alpha } \beta },\nonumber\\
&& \epsilon ^{\alpha  \beta } \left(\eta \sigma \Lambda ^k\right)^{\zeta } \Lambda ^l{}_{\beta },\qquad \Lambda ^2 \eta ^{k l} \eta ^{n o},\qquad \Lambda ^{k l} \eta _{l n} \left(\sigma ^{n o}\right)^{\alpha  \beta },\qquad \sigma \Lambda \Lambda  \eta ^{k l} \epsilon ^{\alpha  \beta },\nonumber\\
&&\eta _{k l} \left(\bar{\sigma }^l\right)^{\dot{\alpha } \beta } \left(\eta \sigma \Lambda \Lambda ^n\right)^k,\qquad \epsilon ^{\alpha  \beta } \eta _{k l} \left(\eta \sigma \Lambda ^l\right)^{\zeta } \Lambda ^k{}_{\beta },\qquad \epsilon ^{\alpha  \beta } \epsilon ^{\zeta  \iota } \epsilon ^{k l n o} \eta _{n p} \eta _{o q} \Lambda ^p{}_{\beta } \Lambda ^q{}_{\iota }.
\end{eqnarray}

\item There exist six different operator factors,
\begin{enumerate}

\item The 1-point function contains only one form,
\begin{equation}
\label{}
D^2 \Phi;
\end{equation}

\item The 2-point function contains ten different forms,
\begin{eqnarray}
\label{}
&&\Phi  \left(D^2 \Phi \right),\qquad
 (D \Phi ) (D \Phi ),\qquad
 \left(D^2 \Phi \right) \left(D^2 \Phi \right),\qquad
 \left(D^2 \Phi \right) \left(\bar{D}^2 \Phi ^+\right),\nonumber\\
&&\left(D^2 \Phi \right) \Phi ^+,\qquad
 \partial (D \Phi ) \left(\bar{D} \Phi ^+\right),\qquad
 \partial \partial \Phi  \Phi ^+,\qquad
 \partial \partial \left(D^2 \Phi \right) \left(D^2 \Phi \right),\nonumber\\
&&\partial \partial \Phi ^+ \Phi ^+,\qquad
 \partial \partial \partial \partial \Phi ^+ \Phi ^+;
\end{eqnarray}

\item The 3-point function contains thirteen different forms,
\begin{eqnarray}
\label{}
&&\Phi  \Phi  \left(D^2 \Phi \right),\qquad
 \Phi  (D \Phi ) (D \Phi ),\qquad
 \Phi  \left(D^2 \Phi \right) \left(D^2 \Phi \right),\qquad
 \Phi  \left(D^2 \Phi \right) \Phi ^+,\nonumber\\
&&(D \Phi ) (D \Phi ) \left(D^2 \Phi \right),\qquad
 (D \Phi ) (D \Phi ) \Phi ^+,\qquad
 \left(D^2 \Phi \right) \left(D^2 \Phi \right) \Phi ^+,
\nonumber\\
&& \left(D^2 \Phi \right) \left(\bar{D} \Phi ^+\right) \left(\bar{D} \Phi ^+\right),\qquad
\left(D^2 \Phi \right) \Phi ^+ \left(\bar{D}^2 \Phi ^+\right),\qquad
 \left(D^2 \Phi \right) \Phi ^+ \Phi ^+, \nonumber\\
&& \partial (D \Phi ) \Phi ^+ \left(\bar{D} \Phi ^+\right),\qquad
\partial \partial \Phi  \Phi ^+ \Phi ^+,\qquad
\partial \partial \Phi ^+ \Phi ^+ \Phi ^+;
\end{eqnarray}

\item The 4-point function contains fourteen different forms,
\begin{eqnarray}
\label{}
&&\Phi  \Phi  \left(D^2 \Phi \right) \left(D^2 \Phi \right),\qquad
 \Phi  \Phi  \left(D^2 \Phi \right) \Phi ^+,\qquad
 \Phi  (D \Phi ) (D \Phi ) \left(D^2 \Phi \right),
\nonumber\\
&& \Phi  (D \Phi ) (D \Phi ) \Phi ^+,\qquad
\Phi  \left(D^2 \Phi \right) \left(D^2 \Phi \right) \Phi ^+,\qquad
 \Phi  \left(D^2 \Phi \right) \Phi ^+ \Phi ^+,
 \nonumber\\
&& (D \Phi ) (D \Phi ) \left(D^2 \Phi \right) \Phi ^+,\qquad
(D \Phi ) (D \Phi ) \Phi ^+ \Phi ^+,\qquad
\left(D^2 \Phi \right) \Phi ^+ \left(\bar{D} \Phi ^+\right) \left(\bar{D} \Phi ^+\right),
\nonumber\\
&&  \left(D^2 \Phi \right) \Phi ^+ \Phi ^+ \left(\bar{D}^2 \Phi ^+\right),\qquad
 \left(D^2 \Phi \right) \Phi ^+ \Phi ^+ \Phi ^+,\qquad
 \partial (D \Phi ) \Phi ^+ \Phi ^+ \left(\bar{D} \Phi ^+\right),
\nonumber\\ 
&&\partial \partial \Phi  \Phi ^+ \Phi ^+ \Phi ^+,\qquad
 \partial \partial \Phi ^+ \Phi ^+ \Phi ^+ \Phi ^+;
\end{eqnarray}

\item The 5-point function contains ten different forms,
\begin{eqnarray}
\label{}
&&\Phi  \Phi  \left(D^2 \Phi \right) \left(D^2 \Phi \right) \Phi ^+,\qquad
 \Phi  \Phi  \left(D^2 \Phi \right) \Phi ^+ \Phi ^+,\qquad
 \Phi  (D \Phi ) (D \Phi ) \left(D^2 \Phi \right) \Phi ^+,\nonumber\\
&& \Phi  (D \Phi ) (D \Phi ) \Phi ^+ \Phi ^+,\qquad
\Phi  \left(D^2 \Phi \right) \Phi ^+ \Phi ^+ \Phi ^+,\qquad
 (D \Phi ) (D \Phi ) \Phi ^+ \Phi ^+ \Phi ^+,\nonumber\\
&& \left(D^2 \Phi \right) \Phi ^+ \Phi ^+ \left(\bar{D} \Phi ^+\right) \left(\bar{D} \Phi ^+\right),\qquad
 \left(D^2 \Phi \right) \Phi ^+ \Phi ^+ \Phi ^+ \left(\bar{D}^2 \Phi ^+\right),\nonumber\\
&&\partial (D \Phi ) \Phi ^+ \Phi ^+ \Phi ^+ \left(\bar{D} \Phi ^+\right),\qquad
 \partial \partial \Phi  \Phi ^+ \Phi ^+ \Phi ^+ \Phi ^+;
\end{eqnarray}

\item The 6-point function contains two different forms,
\begin{eqnarray}
\label{}
\Phi  \Phi  \left(D^2 \Phi \right) \Phi ^+ \Phi ^+ \Phi ^+,\qquad
 \Phi  (D \Phi ) (D \Phi ) \Phi ^+ \Phi ^+ \Phi ^+.
\end{eqnarray}

\end{enumerate}
\end{itemize}

Because of the existence of the above new forms, $\Gamma _{\rm 2nd}\left(\Lambda ^2\right)$ cannot be absorbed by ${\cal S}_{(1)}$, which means ${\cal S}_{(1)}$ is still not renormalizable. Different from ref.~\cite{Grisaru:2003fd}, we have to compute the third successive effective action $\Gamma _{\rm 3rd}$.

\subsection{Formulation and Classification of $\Gamma _{\rm 3rd}$}

\subsubsection{Method}

Similar to the calculation of $\Gamma _{\rm 2nd}\left(\Lambda ^2\right)$, we define
\begin{equation}
\label{}
{\cal S}_{(2)}\equiv {\cal S}_{(1)}+\Gamma _{\rm 2nd}\left(\Lambda ^2\right),
\end{equation}
and calculate the effective action of ${\cal S}_{(2)}$.
Because the effective action contributed by the vertices of  ${\cal S}_{\Lambda }\left(\Lambda ^2\right)+\Gamma _{\rm 1st}\left(\Lambda ^2\right)$ is already known,
what we need to do is just to calculate the effective action of ${\cal S}_{\rm WZ}+\Gamma _{\rm 2nd}\left(\Lambda ^2\right)$.

Different from the calculation of $\Gamma _{\rm 2nd}\left(\Lambda ^2\right)$, $V_{(2)}$ contains the 2-point function of $\Gamma _{\rm 2nd}\left(\Lambda ^2\right)$, and it is determined by ${\cal S}_{{\rm int}}+\Gamma _{\rm 2nd}\left(\Lambda ^2\right)$. Exactly following the way described in subsection 5.6.1, we get the desired result $\Gamma _{\rm 3rd}\left(\Lambda ^2\right)$.

\subsubsection{Result}

By comparing the BFNC parameter factors and operator ones of $\Gamma _{\rm 3rd}\left(\Lambda ^2\right)$ with that of ${\cal S}_{(2)}$, we find the following new BFNC parameter factors that do not appear in ${\cal S}_{(2)}$,
\begin{equation}
\label{}
\eta ^{k l} \left(\eta \sigma \Lambda \Lambda ^n\right)^o,\qquad \sigma \Lambda \Lambda  (\sigma ^{k l})^{\alpha  \beta },
\end{equation}
but the operator factors in $\Gamma _{\rm 3rd}\left(\Lambda ^2\right)$ are exactly same as that in ${\cal S}_{(2)}$.
Therefore, there are still some terms in $\Gamma _{\rm 3rd}\left(\Lambda ^2\right)$ that cannot be absorbed by ${\cal S}_{(2)}$.
The reason is obvious because these terms are the products of the new BFNC parameter factors and the operator factors. This implies that we have to continue  the iterative procedure  in order to find the renormalizable Wess-Zumino model that possesses the $1/2$ supersymmetry invariance on the BFNC superspace. At this stage, we notice that many terms in $\Gamma _{\rm 1st}\left(\Lambda ^2\right)$, $\Gamma _{\rm 2nd}\left(\Lambda ^2\right)$, and $\Gamma _{\rm 3rd}\left(\Lambda ^2\right)$ are same, which brings us to compare the three effective actions and to analyze their structures.

\subsection{$1/2$ Supersymmetry Invariant Subset}

Based on the Wess-Zumino action (eq.~(\ref{WZ action})) and the definition of $\star$-product (eq.~(\ref{star_product_expansion})), ${\cal S}_{\rm NC}$ (eq.~(\ref{S_NC})) is invariant under the $1/2$ supersymmetry transformation (see eq.~(\ref{supersymmetry transformation})).
By using the background field method, we can ensure that all of the effective actions are invariant under the $1/2$ supersymmetry transformation. In ${\cal S}_{\rm NC}$, the noncommutative part ${\cal S}_{\Lambda}$ contains the contributions from the orders of $\Lambda^2$ and $\Lambda^4$, see subsections 5.3.1 and 5.3.2, and each order contribution as a close set is invariant under the $1/2$ supersymmetry transformation.
In the present work, we focus on the order of $\Lambda ^2$ and further find that ${\cal S}_{\Lambda}\left(\Lambda ^2\right)$ can be separated into several classes, each of which as a close set contains the minimal number of terms and is invariant under the $1/2$ supersymmetry transformation. Any two classes have no common terms.
So we introduce the concept of $1/2$ supersymmetry invariant subsets and name every class as a $1/2$ supersymmetry invariant subset.

The number of terms in ${\cal S}_{\Lambda}\left(\Lambda ^2\right)$ is small, so it is easy to find its subsets.
However, each of the effective actions $\Gamma _{\rm 1st}\left(\Lambda ^2\right)$, $\Gamma _{\rm 2nd}\left(\Lambda ^2\right)$, and $\Gamma _{\rm 3rd}\left(\Lambda ^2\right)$ contains a quite large number of terms, and thus a systematic method is needed to find  the $1/2$ supersymmetry invariant subsets for each effective action.
We explain our proposal  as follows by using $\Gamma _{\rm 1st}\left(\Lambda ^2\right)$ as an example.

{\em Step 1}:  After transforming  $\Gamma _{\rm 1st}\left(\Lambda ^2\right)$ to its simplest form, we still have 68 terms which are labelled arbitrarily and denoted as $L_i$,
\begin{equation}
\label{}
\Gamma _{\rm 1st}\left(\Lambda ^2\right)=\sum _{i=1}^{n} {L}_i,
\end{equation}
where $n=68$.

{\em Step 2}: We multiply $L_i$ by a constant $X_i$ and sum them, and denote the result as $A$,
\begin{equation}
\label{}
A=\sum _{i=1}^{n} X_i{L}_i.
\end{equation}

{\em Step 3}: Making the $1/2$ supersymmetry transformation to $A$ and simplifying $\delta _{\xi }A$, we obtain that it consists of $n'$ terms,
\begin{equation}
\label{}
\delta _{\xi }A=\sum _{j=1}^{n'} Y_jL^{\prime}_j,
\end{equation}
where $L^{\prime}_j$ is different from $L_i$. Thus we derive that $Y_j$ is a  linear function of $X_i$'s,
\begin{equation}
\label{}
Y_j=\sum _{i=1}^{n} c_{ji} X_i, 
\end{equation}
where $c_{ji}$ is real. For every $j$, $c_{ji}=0$ for some $i$'s, that is, those $X_i$'s do not appear in $Y_j$.

For each $j$, $j=1, \cdots, n'$, we construct set $U_j$ using the parameters $X_i$'s in $Y_j$, 
\begin{equation}
\label{}
U_j=\{X_{i_1},X_{i_2}, \cdots\},
\end{equation}
where $\{i_1, i_2, \cdots\}$ is a subset of $\{1, \cdots, n\}$.
Then we obtain set $W$ which contains $n'$ elements,
\begin{equation}
\label{}
W=\{U_1,U_2,\cdots,U_{n'}\}.
\end{equation}

{\em Step 4}: For every pair of $(j, j')$, where $j\neq j'$, if $U_{j}\cap U_{j'}\neq\emptyset$, we define
\begin{equation}
\label{}
U_{jj'}\equiv U_{j} \cup U_{j'},
\end{equation}
and remove $U_{j}$ and  $U_{j'}$ from $W$, but add $U_{jj'}$ to $W$.

Repeating the above process,
we obtain at the end,
\begin{equation}
\label{}
W'=\{I_{1},I_{2},\cdots,I_{m},\cdots\},
\end{equation}
where its element
\begin{equation}
\label{}
I_m=\{X_{m_1},X_{m_2},\cdots \}  
\end{equation}
satisfies $I_m\cap I_m'=\emptyset$ when $m\neq m'$, and $\{m_1, m_2, \cdots\}$ is a subset of $\{1, \cdots, n\}$.

{\em Step 5}:  For set $I_m=\{X_{m_1},X_{m_2},\cdots\}$, we sum the elements of its corresponding set $\{L_{m_1},L_{m_2},\cdots\}$, and denote it as $f_m$,
\begin{equation}
\label{}
f_m=\sum _{i=\{m_1, m_2,\cdots\}} L_i.
\end{equation}
Thus, $f_m$ provides one  $1/2$ supersymmetry invariant subset. We  can deduce that there exist 17  invariant subsets in $\Gamma _{\rm 1st}\left(\Lambda ^2\right)$.

By using the above proposal that consists of  five steps, we can separate the other three actions, ${\cal S}_{\Lambda }\left(\Lambda ^2\right)$,  $\Gamma _{\rm 2nd}\left(\Lambda ^2\right)$, and $\Gamma _{\rm 3rd}\left(\Lambda ^2\right)$, into their $1/2$ supersymmetry invariant subsets whose numbers are 4, 64, and 73, respectively. By comparing the invariant subsets of the four actions, we see that some subsets have same BFNC parameter and operator factors but different coefficients. After summing up all of the subsets of the four actions, we finally determine that there are 74 independent subsets in total, each of which is invariant under the $1/2$ supersymmetry transformation, see Appendix B.1 for the details where the explanation of four parameters ${\color{red}a_0}$, ${\color{red}a_1}$, ${\color{red}a_2}$, and ${\color{red}a_3}$ is provided below.

\subsection{Analysis of Effective Actions by Invariant Subsets}

As mentioned in the above that  some terms have same BFNC parameter and operator factors but different coefficients  in the four actions ${\cal S}_{\Lambda }\left(\Lambda ^2\right)$, $\Gamma _{\rm 1st}\left(\Lambda ^2\right)$, $\Gamma _{\rm 2nd}\left(\Lambda ^2\right)$, and $\Gamma _{\rm 3rd}\left(\Lambda ^2\right)$, we present these actions in an explicit way related to invariant subsets, that is,  we show the ingredients of an invariant subset provided by the four actions.

We construct a new action,
\begin{equation}
\label{}
\Gamma\left(\Lambda ^2\right)={\color{red}a_0} \,{\cal S}_{\Lambda }\left(\Lambda ^2\right)
+{\color{red}a_1} \,\Gamma _{\rm 1st}\left(\Lambda ^2\right)
+{\color{red}a_2}\, \Gamma _{\rm 2nd}\left(\Lambda ^2\right)
+{\color{red}a_3}\, \Gamma _{\rm 3rd}\left(\Lambda ^2\right),
\end{equation}
where ${\color{red}a_0}$, ${\color{red}a_1}$, ${\color{red}a_2}$, and ${\color{red}a_3}$ are parameters which show intersections of invariant subsets of different actions.
All invariant subsets for $\Gamma\left(\Lambda ^2\right)$ are given in Appendix~\ref{result_EF} and denoted by $f_i$, where $i=1,\cdots,74$. Thus, $\Gamma\left(\Lambda ^2\right)$ can be rewritten as
\begin{equation}
\label{}
\Gamma\left(\Lambda ^2\right)=\sum_{i=1}^{74} f_i.
\end{equation}

In order to explicitly show the intersections of invariant subsets of different actions, we take the invariant subset No. 29 (see Appendix~\ref{result_EF}), $f_{29}$, as an example,
\begin{eqnarray}
\label{subset_example}
f_{29}&=&\frac{5 \left(-18 g^5 {\color{red} a_2}+23 g^7 {\color{red} a_3}\right)}{13824} \epsilon ^{\alpha  \beta } \Lambda ^{k l} \theta ^4 \left(D_{\beta } \Phi \right) \partial _l\partial _k\left(D_{\alpha } \Phi \right) \left(D^2 \Phi \right)\nonumber\\
&&+\frac{-192 g {\color{red} a_0}-36 g^5 {\color{red} a_2}+55 g^7 {\color{red} a_3}}{3072} \epsilon ^{\alpha  \beta } \epsilon ^{\zeta  \iota } \Lambda ^k{}_{\beta } \Lambda ^l{}_{\iota } \theta ^4 \partial _k\left(D_{\alpha } \Phi \right) \partial _l\left(D_{\zeta } \Phi \right) \left(D^2 \Phi \right)\nonumber\\
&&+\frac{-144 g {\color{red} a_0}-45 g^5 {\color{red} a_2}+68 g^7 {\color{red} a_3}}{4608} \Lambda ^{k l} \theta ^4 \Phi  \partial _k\left(D^2 \Phi \right) \partial _l\left(D^2 \Phi \right)\nonumber\\
&&+\frac{-864 g {\color{red} a_0}-360 g^5 {\color{red} a_2}+523 g^7 {\color{red} a_3}}{27648} \Lambda ^{k l} \theta ^4 \Phi  \left(D^2 \Phi \right) \partial _l\partial _k\left(D^2 \Phi \right)\nonumber\\
&&+\frac{-3456 g {\color{red} a_0}-504 g^5 {\color{red} a_2}+851 g^7 {\color{red} a_3}}{55296} \epsilon ^{\alpha  \beta } \Lambda ^{k l} \theta ^4 \partial _k\left(D_{\alpha } \Phi \right) \partial _l\left(D_{\beta } \Phi \right) \left(D^2 \Phi \right).
\end{eqnarray}
The first line means that the combination of BFNC parameter and operator factors, $$\epsilon ^{\alpha  \beta } \Lambda ^{k l} \theta ^4 \left(D_{\beta } \Phi \right) \partial _l\partial _k\left(D_{\alpha } \Phi \right) \left(D^2 \Phi \right),$$ comes from both $\Gamma _{\rm 2nd}\left(\Lambda ^2\right)$ and $\Gamma _{\rm 3rd}\left(\Lambda ^2\right)$, but the two effective actions provide different coefficients which are $\frac{-90 g^5}{13824}$ and $\frac{115 g^7}{13824}$, respectively.  
The other four lines have the similar meaning, and emerge from ${\cal S}_{\Lambda }\left(\Lambda ^2\right)$, $\Gamma _{\rm 2nd}\left(\Lambda ^2\right)$, and $\Gamma _{\rm 3rd}\left(\Lambda ^2\right)$ because  ${\color{red}a_0}$, ${\color{red}a_2}$, and ${\color{red}a_3}$ appear. Moreover, $\Gamma _{\rm 1st}\left(\Lambda ^2\right)$ has no contribution  to $f_{29}$ as ${\color{red}a_1}$ does not appear. In this way, one can clearly see the ingredients of every invariant subset provided by ${\cal S}_{\Lambda }\left(\Lambda ^2\right)$, $\Gamma _{\rm 1st}\left(\Lambda ^2\right)$, $\Gamma _{\rm 2nd}\left(\Lambda ^2\right)$, and $\Gamma _{\rm 3rd}\left(\Lambda ^2\right)$.

\subsection{Basis of Supersymmetry Invariant Subset}

Based on the analysis made to the 74 supersymmetry invariant subsets, we try to construct more general $1/2$ supersymmetry invariant subsets in order to deduce the one-loop renormalizable Wess-Zumino action on the BFNC superspace. For the sake of concreteness, we take $f_{29}$ as an example for proposing our idea. $f_{29}$ has five different combinations of BFNC parameter and operator factors, in each combination 
the coefficients from ${\cal S}_{\Lambda }\left(\Lambda ^2\right)$, $\Gamma _{\rm 2nd}\left(\Lambda ^2\right)$, and $\Gamma _{\rm 3rd}\left(\Lambda ^2\right)$ are different.  
Four terms corresponding to $\color{red} a_0$ are from ${\cal S}_{\Lambda }\left(\Lambda ^2\right)$, five terms corresponding to $\color{red} a_2$ are from $\Gamma _{\rm 2nd}\left(\Lambda ^2\right)$, and the last five terms corresponding to $\color{red} a_3$ are from $\Gamma _{\rm 3rd}\left(\Lambda ^2\right)$. We have already known that the three groups of terms possess the $1/2$ supersymmetry invariance because they belong to the invariant subsets of ${\cal S}_{\Lambda }\left(\Lambda ^2\right)$,  $\Gamma _{\rm 2nd}\left(\Lambda ^2\right)$, and $\Gamma _{\rm 3rd}\left(\Lambda ^2\right)$, respectively.
To construct a more general $1/2$ supersymmetry invariant subset than $f_{29}$, we try to multiply every combination by an arbitrary parameter, then look for the constraints of the parameters by demanding that the generalization of $f_{29}$ is $1/2$ supersymmetry invariant, and at last give the generalization that contains arbitrary parameters (other than $\color{red} a_0$, $\color{red} a_2$, and $\color{red} a_3$). 
In this way, we introduce the concept of a basis for every $1/2$ supersymmetry invariant subset, meaning a generalized invariant subset with arbitrary parameters. 
We describe our proposal in more details below on how to work out a basis of a $1/2$ supersymmetry invariant subset.

We write an invariant subset $f_i$ as
\begin{equation}
\label{}
f_i=\sum _{j=1}^{n_i}C_{ij}L_j, 
\end{equation}
where $C_{ij}$ 
stands for a coefficient and $L_j$ a combination of BFNC parameter and operator factors, $i=1, 2, \cdots, 74$ the total number of invariant subsets, and $n_i$ means the number of terms in $f_i$.

{\em Step 1}: By introducing three groups of parameters ${\color{blue}x_{i,j}}$, ${\color{blue}y_{i,j}}$, and ${\color{blue}z_{i,j}}$, we construct a basis $A_i$,
\begin{equation}
\label{}
A_i\equiv\sum _{j=1}^{n_i} \left({\color{blue}x_{i,j}}+{\color{blue}y_{i,j}}\Lambda ^2+{\color{blue}z_{i,j}}\sigma \Lambda \Lambda \right)L_j.
\end{equation}

{\em Step 2}: Remove from $A_i$ those terms where the power of $\Lambda$ is greater than 2 because $L_j$ contains $\Lambda^2$ and $\sigma\Lambda\Lambda$ in its BFNC parameter factors. 
Then some components of ${\color{blue}x_{i,j}}$, ${\color{blue}y_{i,j}}$, and ${\color{blue}z_{i,j}}$ can be eliminated.
We denote the result as $A^{\prime}_i$.

{\em Step 3}: Making the $1/2$ supersymmetry transformation for $A^{\prime}_i$, we obtain
\begin{equation}
\label{}
\delta A^{\prime}_i=\sum _{k=1}^{n^{\prime}_i} G_{i,k}(x,y,z)L^{\prime}_k,
\end{equation}
where $G_{i,k}(x,y,z)$ is a linear function of ${\color{blue}x_{i,j}}$, ${\color{blue}y_{i,j}}$, and ${\color{blue}z_{i,j}}$, where $j=1,...,n_i$.

{\em Step 4}: Let $A^{\prime}_i$ be a $1/2$ invariant subset, we have $\delta A^{\prime}_i=0$. 
Because $L^{\prime}_k$'s are independent to each other, we get
\begin{equation}
\label{equation_G}
G_{i,k}(x,y,z)=0,
\end{equation}
which is a set of constraints on parameters ${\color{blue}x_{i,j}}$, ${\color{blue}y_{i,j}}$, and ${\color{blue}z_{i,j}}$.

{\em Step 5}: Not all of  ${\color{blue}x_{i,j}}$, ${\color{blue}y_{i,j}}$, and ${\color{blue}z_{i,j}}$ are independent. 
We eliminate the dependent parameters from $A^{\prime}_i$ and then obtain a basis.

As an example, we explain the basis $B_{29}$ that corresponds to the subset $f_{29}$ (eq.~(\ref{subset_example})).
By using the above proposal, we obtain
\begin{eqnarray}
\label{basis_example}
B_{29}&=&\left(-2 {\color{blue}x_{29,3}}+2 {\color{blue}x_{29,4}}\right) \epsilon ^{\alpha  \beta } \Lambda ^{k l} \theta ^4 \left(D_{\beta } \Phi \right) \partial _l\partial _k\left(D_{\alpha } \Phi \right) \left(D^2 \Phi \right)\nonumber\\
&&+\left(-6 {\color{blue}x_{29,3}}+4 {\color{blue}x_{29,4}}+2 {\color{blue}x_{29,5}}\right) \epsilon ^{\alpha  \beta } \epsilon ^{\zeta  \iota } \Lambda ^k{}_{\beta } \Lambda ^l{}_{\iota } \theta ^4 \partial _k\left(D_{\alpha } \Phi \right) \partial _l\left(D_{\zeta } \Phi \right) \left(D^2 \Phi \right)\nonumber\\
&&+{\color{blue}x_{29,3}} \Lambda ^{k l} \theta ^4 \Phi  \partial _k\left(D^2 \Phi \right) \partial _l\left(D^2 \Phi \right)\nonumber\\
&&+{\color{blue}x_{29,4}} \Lambda ^{k l} \theta ^4 \Phi  \left(D^2 \Phi \right) \partial _l\partial _k\left(D^2 \Phi \right)\nonumber\\
&&+{\color{blue}x_{29,5}} \epsilon ^{\alpha  \beta } \Lambda ^{k l} \theta ^4 \partial _k\left(D_{\alpha } \Phi \right) \partial _l\left(D_{\beta } \Phi \right) \left(D^2 \Phi \right),
\end{eqnarray}
where ${\color{blue}x_{29,3}}$, ${\color{blue}x_{29,4}}$, and ${\color{blue}x_{29,5}}$ are independent parameters that are related to lines $3$, $4$, and $5$ of $f_{29}$.
The parameters ${\color{blue}x_{29,1}}$ and ${\color{blue}x_{29,2}}$ related to lines $1$ and $2$ of $f_{29}$ are not independent but determined by the three independent parameters, i.e., ${\color{blue}x_{29,1}}=-2 {\color{blue}x_{29,3}}+2 {\color{blue}x_{29,4}}$ and ${\color{blue}x_{29,2}}=-6 {\color{blue}x_{29,3}}+4 {\color{blue}x_{29,4}}+2 {\color{blue}x_{29,5}}$.

We observe that in subset $f_{29}$  the terms that belong to actions ${\cal S}_{\Lambda }\left(\Lambda ^2\right)$, $\Gamma _{\rm 2nd}\left(\Lambda ^2\right)$, and $\Gamma _{\rm 3rd}\left(\Lambda ^2\right)$, respectively, can be obtained by choosing special values for the independent parameters ${\color{blue}x_{29,3}}$, ${\color{blue}x_{29,4}}$, and ${\color{blue}x_{29,5}}$ in $B_{29}$ (eq.~(\ref{basis_example})). 
We list the corresponding values for ${\cal S}_{\Lambda }\left(\Lambda ^2\right)$, $\Gamma _{\rm 2nd}\left(\Lambda ^2\right)$, or $\Gamma _{\rm 3rd}\left(\Lambda ^2\right)$ in the following three lines, respectively, 
\begin{eqnarray}
\label{parameters_example}
&&{\color{blue}x_{29,3}}\to -\frac{1}{32},\qquad
{\color{blue}x_{29,4}}\to -\frac{1}{32},\qquad
{\color{blue}x_{29,5}}\to -\frac{1}{16};\nonumber\\
&&{\color{blue}x_{29,3}}\to -\frac{5}{512},\qquad
{\color{blue}x_{29,4}}\to -\frac{5}{384},\qquad
{\color{blue}x_{29,5}}\to -\frac{7}{768};\nonumber\\
&&{\color{blue}x_{29,3}}\to \frac{17}{1152},\qquad
{\color{blue}x_{29,4}}\to \frac{523}{27648},\qquad
{\color{blue}x_{29,5}}\to \frac{851}{55296}.
\end{eqnarray}
This means
if we replace the parameters ${\color{blue}x_{29,3}}$, ${\color{blue}x_{29,4}}$, and ${\color{blue}x_{29,5}}$ in $B_{29}$ by the values of the first line of eq.~(\ref{parameters_example}), we obtain the contribution from ${\cal S}_{\Lambda }\left(\Lambda ^2\right)$ to the subset $f_{29}$, which is equivalently to set ${\color{red}a_0}=1$, ${\color{red}a_2}=0$, and ${\color{red}a_3}=0$ in $f_{29}$ (eq.~(\ref{subset_example})).

We deal with every subset $f_i$ using our proposal, and obtain its corresponding basis $B_i$, where $i=1,\cdots,74$. That is, there are 74 bases in total that are given in Appendix~\ref{susy_base}. By using the bases, we shall get the desired renormalizable action.

\subsection{Renormalizable Action}

According to the background field method, we need to calculate the effective action $\Gamma _{\rm 4th}\left(\Lambda ^2\right)$ based on  ${\cal S}_{\rm WZ}+{\cal S}_{\Lambda }\left(\Lambda ^2\right)+ \Gamma _{\rm 1st}\left(\Lambda ^2\right)+ \Gamma _{\rm 2nd}\left(\Lambda ^2\right)+\Gamma _{\rm 3rd}\left(\Lambda ^2\right)$. From the above discussions we discover
that it is more convenient to use the bases $B_i$'s rather than  the  four actions ${\cal S}_{\Lambda }\left(\Lambda ^2\right)$, $\Gamma _{\rm 1st}\left(\Lambda ^2\right)$, $\Gamma _{\rm 2nd}\left(\Lambda ^2\right)$, and $\Gamma _{\rm 3rd}\left(\Lambda ^2\right)$. Therefore, we construct ${\cal S}_{(3)}$ as follows,
\begin{equation}
\label{S_3}
{\cal S}_{(3)}={\cal S}_{\rm WZ}+\int d^8z\left(\sum _{i=1}^{74} B_i\right),
\end{equation}
whose effective action is just $\Gamma _{\rm 4th}\left(\Lambda ^2\right)$. Following the method adopted in subsections 5.6 and 5.7, we work out $\Gamma _{\rm 4th}\left(\Lambda ^2\right)$ and observe\footnote{Because $\Gamma _{\rm 4th}\left(\Lambda ^2\right)$ takes the similar form to ${\cal S}_{(3)}$, i.e., its combinations of BFNC parameter and operator factors are covered by the 74 invariant bases $B_i$'s and the only difference from ${\cal S}_{(3)}$ is in its coefficients, so it is not necessary for us to write it explicitly.} that it no longer has new terms that do not exist in ${\cal S}_{(3)}$, i.e.,  $\Gamma _{\rm 4th}\left(\Lambda ^2\right)$ can be absorbed completely by ${\cal S}_{(3)}$.

In the process of calculation, although we do not give the Feynman graphs for each term in the effective action, we can obtain the information about where the terms of the effective action come from by calculating the supertrace of matrix. The action ${\cal S}_{(3)}$ contains parameters ${\color{blue}x_{i,j}}$, ${\color{blue}y_{i,j}}$, and ${\color{blue}z_{i,j}}$. In the effective action of ${\cal S}_{(3)}$, the coefficients of each term are functions of the parameters. From these parameters we can determine where the terms of the effective action come from. On the BFNC superspace we conclude that this method is more convenient than Feynman rules for making a deformed action renormalizable. 

We note that the divergence of each term in the effective action is obvious because each divergent term contains the factor $1/\epsilon$, which can be seen when the momentum integral is evaluated by using the dimensional regulation and minimal subtraction method. What we want to emphasize in our paper is how to add new terms to make the deformed model renormalizable.

Now we discuss if the renormalization of parameters is compatible. 
The bare action is given by~\cite{Dimitrijevic:2010yv},
\begin{equation}
\label{}
{\cal S}_{\rm B}={\cal S}_{(3)}-\Gamma _1\bigg|_{{\rm dp}},
\end{equation}
where $\Gamma _1|_{{\rm dp}}$ equals to the effective action $\Gamma_{\rm 4th}\left(\Lambda ^2\right)$ we just mentioned.

The parameters in ${\cal S}_{(3)}$ include $m$ and $g$ provided by action ${\cal S}_{\rm WZ}$ (eq.~(\ref{WZ_superfield_form})) and ${\color{blue}x_{i,j}}$, ${\color{blue}y_{i,j}}$, and ${\color{blue}z_{i,j}}$ provided by the $1/2$ supersymmetry invariant bases (eq.~(\ref{susy_base})).

We at first analyze  the renormalization of fields. The following term exists in $\Gamma_{\rm 4th}\left(\Lambda ^2\right)$,
\begin{equation}
\label{}
\frac{g^2}{4\pi ^2\epsilon }\Phi ^+\Phi,
\end{equation}
so the corresponding term in ${\cal S}_{\rm B}$ is
\begin{equation}
\label{field_2_point}
\left(1-\frac{g^2}{4\pi ^2\epsilon }\right)\Phi ^+\Phi.
\end{equation}
The basic idea of renormalization is to redefine fields and parameters in order 
to transform action ${\cal S}_{\rm B}$ to the form of ${\cal S}_{(3)}$. 
We introduce $Z$ for field renormalization,
\begin{equation}
\label{define_Z}
Z=\left(1-\frac{g^2}{4\pi ^2\epsilon }\right),
\end{equation}
then according to eqs.~(\ref{field_2_point}) and (\ref{define_Z}) redefine $\Phi$ and $\Phi ^+$ by
$Z \Phi ^+\Phi \equiv \Phi _0^+\Phi _0$. By this way we have
\begin{equation}
\label{field_renormalization}
\sqrt{Z}\Phi ^+= \Phi _0{}^+,\qquad \sqrt{Z}\Phi = \Phi _0,
\end{equation}
which is just same as the renormalization of fields for the ordinary Wess-Zumino model.

Now we study the renormalization of mass $m$. Because the following terms are not contained in $\Gamma_{\rm 4th}\left(\Lambda ^2\right)$,
\begin{equation}
\label{}
\Phi \left(\frac{D^2}{\square }\Phi \right),\qquad \Phi ^+\left(\frac{\bar{D}^2}{\square }\Phi ^+\right),
\end{equation}
the renormalization of $m$ is completely determined by the renormalization of fields. By using eq.~(\ref{field_renormalization}), we have
\begin{equation}
\label{mass_renormalization}
m\Phi \left(\frac{D^2}{\square }\Phi \right)=m\left(\frac{1}{\sqrt{Z}}\right)^2 \Phi_0 \left(\frac{D^2}{\square }\Phi_0 \right).
\end{equation}
We just need to define
\begin{equation}
\label{redefine_m}
m\left(\frac{1}{\sqrt{Z}}\right)^2 \equiv m_0,
\end{equation}
then eq.~(\ref{mass_renormalization}) changes to be
\begin{equation}
\label{}
m \Phi \left(\frac{D^2}{\square }\Phi \right)=m_0 \Phi_0 \left(\frac{D^2}{\square }\Phi_0 \right).
\end{equation}
Eq.~(\ref{redefine_m}) is the renormalization of $m$.

In light of eq.~(\ref{field_renormalization}), for $m\Phi ^+\left(\frac{\bar{D}^2}{\square }\Phi ^+\right)$ we have 
\begin{equation}
\label{mass_renormalization_1}
m \Phi ^+\left(\frac{\bar{D}^2}{\square }\Phi ^+\right)=m\left(\frac{1}{\sqrt{Z}}\right)^2 \Phi_0{}^+\left(\frac{\bar{D}^2}{\square }\Phi_0{}^+\right).
\end{equation}
In terms of the renormalization of $m$ in eq.~(\ref{redefine_m}), we transform eq.~(\ref{mass_renormalization_1}) as follows,
\begin{equation}
\label{}
m \Phi ^+\left(\frac{\bar{D}^2}{\square }\Phi ^+\right)=m_0 \Phi_0{}^+\left(\frac{\bar{D}^2}{\square }\Phi_0{}^+\right).
\end{equation}
Therefore, we conclude that for the following two terms,
\begin{equation}
\label{}
m\Phi \left(\frac{D^2}{\square }\Phi \right),\qquad m\Phi ^+\left(\frac{\bar{D}^2}{\square }\Phi ^+\right),
\end{equation}
the renormalization of $m$ is same. In other words, we can say that the renormalizations of $m$ for these two terms are compatible.

Next, we investigate the renormalization of interacting parameter $g$. The following terms are not contained in $\Gamma_{\rm 4th}\left(\Lambda ^2\right)$,
\begin{equation}
\label{}
\Phi  \Phi \left(\frac{D^2}{\square }\Phi \right),\qquad \Phi ^+\Phi ^+\left(\frac{\bar{D}^2}{\square }\Phi ^+\right),
\end{equation}
so the renormalization of $g$ is completely determined by the renormalization of fields, too. By using eq.~(\ref{field_renormalization}), we have
\begin{equation}
\label{g_renormalization}
g \Phi  \Phi \left(\frac{D^2}{\square }\Phi \right)=g\left(\frac{1}{\sqrt{Z}}\right)^3 \Phi _0 \Phi _0\left(\frac{D^2}{\square }\Phi _0\right).
\end{equation}
We just need to define
\begin{equation}
\label{redefine_g}
g\left(\frac{1}{\sqrt{Z}}\right)^3 \equiv g_0,
\end{equation}
then eq.~(\ref{g_renormalization})  changes to be
\begin{equation}
\label{}
g \Phi\Phi \left(\frac{D^2}{\square }\Phi \right)=g_0 \Phi_0\Phi_0 \left(\frac{D^2}{\square }\Phi_0 \right).
\end{equation}
Eq.~(\ref{redefine_g}) is the renormalization of $g$.
The treatment is same for $g\Phi ^+\Phi ^+\left(\frac{\bar{D}^2}{\square }\Phi ^+\right)$. In consequence, we conclude that for the following two terms,
\begin{equation}
\label{}
g\Phi\Phi \left(\frac{D^2}{\square }\Phi \right),\qquad g\Phi ^+\Phi ^+\left(\frac{\bar{D}^2}{\square }\Phi ^+\right),
\end{equation}
the renormalization of $g$ is compatible.

At last we discuss the renormalization of parameters ${\color{blue}x_{i,j}}$, ${\color{blue}y_{i,j}}$, and ${\color{blue}z_{i,j}}$ that appear in the  $1/2$ supersymmetry invariant bases. We take $B_{29}$ (eq.~(\ref{basis_example})) as an example. The following terms appear in the effective action $\Gamma_{\rm 4th}\left(\Lambda ^2\right)$,
\begin{eqnarray}
\label{gamma_4th_B_29}
B^{\prime}_{29}&=&\left(-2 x^{\prime}_{29,3}+2 x^{\prime}_{29,4}\right) \epsilon ^{\alpha  \beta } \Lambda ^{k l} \theta ^4 \left(D_{\beta } \Phi \right) \partial _l\partial _k\left(D_{\alpha } \Phi \right) \left(D^2 \Phi \right)\nonumber\\
&&+\left(-6 x^{\prime}_{29,3}+4 x^{\prime}_{29,4}+2 x^{\prime}_{29,5}\right) \epsilon ^{\alpha  \beta } \epsilon ^{\zeta  \iota } \Lambda ^k{}_{\beta } \Lambda ^l{}_{\iota } \theta ^4 \partial _k\left(D_{\alpha } \Phi \right) \partial _l\left(D_{\zeta } \Phi \right) \left(D^2 \Phi \right)\nonumber\\
&&+x^{\prime}_{29,3} \Lambda ^{k l} \theta ^4 \Phi  \partial _k\left(D^2 \Phi \right) \partial _l\left(D^2 \Phi \right)\nonumber\\
&&+x^{\prime}_{29,4} \Lambda ^{k l} \theta ^4 \Phi  \left(D^2 \Phi \right) \partial _l\partial _k\left(D^2 \Phi \right)\nonumber\\
&&+x^{\prime}_{29,5} \epsilon ^{\alpha  \beta } \Lambda ^{k l} \theta ^4 \partial _k\left(D_{\alpha } \Phi \right) \partial _l\left(D_{\beta } \Phi \right) \left(D^2 \Phi \right),
\end{eqnarray}
where the coefficients $x^{\prime}_{29,3}$, $x^{\prime}_{29,4}$, and $x^{\prime}_{29,5}$ are linear functions of ${\color{blue}x_{i,j}}$'s
and take the following forms,
\begin{eqnarray}
\label{}
x'_{29,3}&=&\frac{-g }{32 \pi ^2 \epsilon }\left(4 x_{21,23}-x_{21,26}+2 x_{21,27}\right),\nonumber\\
x'_{29,4}&=&\frac{-g }{96 \pi ^2 \epsilon }\left(2 x_{20,2}+6 x_{21,19}+12 x_{21,23}-3 x_{21,26}+6 x_{21,27}\right),\nonumber\\
x'_{29,5}&=&\frac{g }{48 \pi ^2 \epsilon }\left(2 x_{20,2}+6 x_{21,19}-6 x_{21,23}-3 x_{21,26}-3 x_{21,27}\right).\nonumber
\end{eqnarray}
One can see that $x'_{29,3}$ depends on the parameters $x_{21,23}$, $x_{21,26}$, and $x_{21,27}$ of $B_{21}$, and $x'_{29,4}$ and $x'_{29,5}$ depend on the parameter $x_{20,2}$ of $B_{20}$ and the parameters $x_{21,19}$, $x_{21,23}$, $x_{21,26}$, and $x_{21,27}$ of $B_{21}$.

For the third line of eq.~(\ref{gamma_4th_B_29}), the corresponding terms in $S_B$ are
\begin{eqnarray}
\label{line_3_in_S_B}
&&{\color{blue}x_{29,3}}\Lambda ^{{kl}}\theta ^4\Phi \partial _k\left(D^2\Phi \right)\partial _l\left(D^2\Phi \right)-x^{\prime}_{29,3}\Lambda ^{{kl}}\theta ^4\Phi \partial _k\left(D^2\Phi \right)\partial _l\left(D^2\Phi \right)\nonumber\\
&=&\left({\color{blue}x_{29,3}}-x^{\prime}_{29,3}\right)\Lambda ^{{kl}}\theta ^4\Phi \partial _k\left(D^2\Phi \right)\partial _l\left(D^2\Phi \right)\nonumber\\
&=&\left({\color{blue}x_{29,3}}-x^{\prime}_{29,3}\right){\Lambda^{{kl}}}\theta ^4\frac{\Phi _0}{\sqrt{Z}}\partial _k\left(D^2\frac{\Phi _0}{\sqrt{Z}}\right)\partial _l\left(D^2\frac{\Phi _0}{\sqrt{Z}}\right)\nonumber\\
&=&{\color{blue}x_{29,3}}\left(1-\frac{x^{\prime}_{29,3}}{{\color{blue}x_{29,3}}}\right)\left(\frac{1}{\sqrt{Z}}\right)^3\Lambda^{{kl}}\theta ^4\Phi _0\partial _k\left(D^2\Phi _0\right)\partial _l\left(D^2\Phi _0\right),
\end{eqnarray}
where we do not need to define the renormalization of $\Lambda ^{{kl}}$.
For analyzing renormalization we have to redefine ${\color{blue}x_{29,3}}$ as $\left(x_0\right)_{29,3}$ in order to transform eq.~(\ref{line_3_in_S_B}) into the following form,
\begin{equation}
\label{r_condition}
\left(x_0\right)_{29,3}\Lambda^{{kl}}\theta ^4\Phi _0\partial _k\left(D^2\Phi _0\right)\partial _l\left(D^2\Phi _0\right).
\end{equation}
By using eqs.~(\ref{line_3_in_S_B}) and (\ref{r_condition}), we obtain
\begin{equation}
\label{x_29_3_redefine}
\left(x_0\right)_{29,3}={\color{blue}x_{29,3}}\left(1-\frac{x^{\prime}_{29,3}}{{\color{blue}x_{29,3}}}\right)\left(\frac{1}{\sqrt{Z}}\right)^3.
\end{equation}
This is the renormalization of parameter ${\color{blue}x_{29,3}}$.

Because both $B_{29}$ and $B^{\prime}_{29}$ are $1/2$ supersymmetry invariant bases and the numbers of their independent parameters are same, the renormalization of their parameters is compatible. The similar discussion can be applied to all of the parameters ${\color{blue}x_{i,j}}$, ${\color{blue}y_{i,j}}$, and ${\color{blue}z_{i,j}}$ of the $1/2$ supersymmetry invariant bases. As a result, we prove that the renormalization of all the parameter $m$, $g$, ${\color{blue}x_{i,j}}$, ${\color{blue}y_{i,j}}$, and ${\color{blue}z_{i,j}}$ is compatible.\footnote{We note that in ref.~\cite{Dimitrijevic:2010yv} the  same parameters are used  in different terms of actions. Such a treatment is doubtful since there no symmetries ensure that the parameters are compatible for renormalization.}

In consequence, we can conclude that the action ${\cal S}_{(3)}$ is one-loop renormalizable.

In the following we try to give some mathematical and/or physical meaning of the constitutive terms in ${\cal S}_{(3)}$.

Normally there are two approaches for deriving a one-loop renormalizable action.

The first is based on the iteration method. One can  calculate an effective action of a deformed action, and then promote the effective action to consist of  1/2 supersymmetry invariant subsets.

The second is based on the symmetry analysis method, see, for instance, our recent work~\cite{Miao:2014b}. One can take into account NC parameters and $D$ algebraic relations in order to obtain general divergent operators. In our case, by combining the BFNC parameters with the operators in all possible ways, we thus deduce all necessary correction terms. By multiplying arbitrary parameters to each term we derive constraints corresponding to the 1/2 supersymmetry invariance. By solving these constraints, we find that some parameters must be zero or depend on other parameters. In this way, we can provide all of the divergent terms in the effective action, and construct the renormalizable action to all order loop expansion using these terms to a large part of which the one-loop renormalizable action corresponds.

In this paper we adopt the first approach. Because we use the background field method and calculate the effective action by using the iteration method, this process involves almost all of the combinations of BFNC parameters and divergent operators. From these combinations we obtain the 1/2 supersymmetry invariant subsets. As there are many possible combinations for certain BFNC parameters and divergent operators, the corresponding invariant subsets naturally contain many correction terms.

In the Wess-Zumino model, because of the supersymmetry, if we represent the action in terms of component fields, the coefficients of different terms are related to each other. For the similar reason, due to  the 1/2 supersymmetry, there exist relations between coefficients of different terms in the subsets $B_i$.

On the BFNC superspace, if we start form the Wess-Zumino model and replace the ordinary product by the BFNC star product, in order to construct a renormalizable action, we must include all of the divergent operators. These terms can be interpreted as the effect of non-locality of the BFNC star product on the physical model. Because the noncommutativity of superspace modifies the quantum properties of the model defined on it, only with these new terms can we make the model renormalizable.

\section{Conclusion and Outlook}
\label{Conclusion_and_Outlook}

Through analyzing  the Wess-Zumino model on the BFNC superspace, we know that the action obtained by replacing the ordinary product by the star product is not renormalizable in general. To make it renormalizable one should add to it correction terms. For the NAC superspace, which is a simpler case, one just needs to add the terms from the primary one-loop effective action, 
and then provides the renormalizable action to all orders 
in perturbation theory~\cite{Britto:2003kg,Romagnoni:2003xt}. 
However, for the BFNC superspace the situation is much more complicated. The iterative process should go up to the third time. Moreover, the complexity also includes that the obtained renormalizable action has so many terms that can be classified on the one hand into 74 subsets each of which has the $1/2$ supersymmetry invariance, and on the other hand into 74 bases that correspond to the 74 subsets.  In particular, in light of the invariant bases 
we construct the one-loop renormalizable action (eq.~(\ref{S_3})) up to the second order of the BFNC parameters $\Lambda ^{k \alpha }$'s.  

In this paper we restrict our discussion to the second order approximation of the BFNC parameters. On the one hand, the renormalization of the model is independent of the orders of the BFNC parameters. 
On the other hand, it gives enough information about the renormalization properties of the model to deal with the second order approximation of the BFNC parameters. We need to calculate the effective action for three times, and in each time we have to manage hundreds of terms. Although the iteration method is not easy to be manipulated, we indeed finish the calculation at the second order of the BFNC parameters. 

For a higher order approximation, we can also process by using the iteration method. Of course, the actual calculation is very complicated. However, in our revet work~\cite{Miao:2014b}, we construct the all order loop renormalizable Wess-Zumino model on the BFNC superspace in the second order approximation of the BFNC parameters. What we have learned can be applied to the analysis of higher order approximations of the BFNC parameters. At the second order approximation of the BFNC parameters, the terms in the all order loop renormalizable Wess-Zumino model are the most general 1/2 supersymmetry invariant forms that originate from the combinations of the BFNC parameters and divergent operators. These forms constitute 1/2 supersymmetry invariant subsets. 

From this observation, we provide another method to obtain the all order loop renormalizable Wess-Zumino model in the second order approximation of the BFNC parameters. We do not need to use the iteration method to calculate the effective action. We just need to construct the most general divergent operators and then combine them with all of the possible factors of the BFNC parameters. Next, we  construct 1/2 supersymmetry invariant subsets and the bases. At the end we obtain the all order loop renormalizable Wess-Zumino model in the second order approximation of the BFNC parameters.

This method can be applied to the higher order approximations of the BFNC parameters, which can avoid the use of the iteration method. We convert the problem into the construction of all possible factors of higher order approximations of the BFNC parameters. In fact, they are finite with orders 2, 4, 6, and 8 of the BFNC parameters. Consequently, the analysis at higher order approximations of the BFNC parameters can be handled in principle. We shall consider the issue in our future work. 
In addition, based on the all order loop renormalizable Wess-Zumino model in the second order approximation of the BFNC parameters~\cite{Miao:2014b}, we shall also extend the corresponding  analysis to orders 4, 6, and 8 of the BFNC parameters $\Lambda ^{k}{}_{ \alpha }$'s.

 
Moreover, the super Yang-Mills in NAC superspace has been constructed in ref.~\cite{Araki:2003se,Lunin:2003bm,Penati:2004gh,Grisaru:2005we,Penati:2009sw,Bianchi:2009fn}, we shall explore the super Yang-Mills model on the BFNC superspace and search for its renormalizable formulation by following the way we have utilized for the Wess-Zumino model.

It will help us in understanding the geometric and physical meaning of the correction terms to consider their bosonic limit and to study the quantum-mechanical reduction of the model. We shall consider these issues in our future work.

\section*{Acknowledgments}
This work was supported in part by the National Natural Science Foundation of China under grant No.11175090 and by the Ministry of Education of China under grant No.20120031110027.
At last, the authors would like to thank the anonymous referees for their helpful comments that indeed improve this work greatly.

\newpage

\section*{Appendix}

\appendix

\setcounter{equation}{0}
\renewcommand\theequation{A\arabic{equation}}

\section{Algebraic Relations}

The following identities of $\sigma$ algebra are used in our calculations of effective actions.
{\small
\begin{eqnarray}
\label{identities_used}
\epsilon ^{k l m n} \eta _{k l}&=&0,\qquad \theta ^{\alpha } \theta ^{\beta } \theta ^{\gamma }=0,\nonumber\\
\left(\bar{\sigma }^{k l}\right)^{\dot{\alpha }}{}_{\dot{\alpha }}&=&0,\qquad \left(\sigma ^{k l}\right)_{\alpha }^{\alpha }=0,\qquad \left(\bar{\sigma }^{k k}\right)^{\dot{\alpha }}{}_{\dot{\beta }}=0,\qquad \left(\sigma ^{k k}\right)_{\alpha }^{\beta }=0,\nonumber\\
\delta _{\alpha  \alpha }&=&2 ,\qquad \delta _{k k}=4 ,\qquad \eta ^{k l} \eta _{l m}=\delta _{k m},\qquad \epsilon ^{\alpha  \beta } \epsilon _{\beta  \gamma }=\delta _{\alpha  \gamma },\nonumber\\
\theta _{\alpha } \theta _{\beta }&=&\frac{1}{2} \epsilon _{\alpha  \beta } \theta ^{\zeta } \theta _{\zeta },\qquad \bar{\theta }_{\dot{\gamma }} \bar{\theta }_{\dot{\kappa }}=-\frac{1}{2} \epsilon _{\dot{\gamma } \dot{\kappa }} \bar{\theta }_{\dot{\alpha }} \bar{\theta }^{\dot{\alpha }},\nonumber\\
\sigma ^l{}_{\beta  \dot{\gamma }} \left(\bar{\sigma }^k\right)^{\dot{\alpha } \beta }&=&-\delta _{\dot{\alpha } \dot{\gamma }} \eta ^{k l}+2 \left(\bar{\sigma }^{k l}\right)^{\dot{\alpha }}{}_{\dot{\gamma }},\nonumber\\
\sigma ^k{}_{\alpha  \dot{\beta }} \left(\bar{\sigma }^l\right)^{\dot{\beta } \gamma }&=&-\delta _{\alpha  \gamma } \eta ^{k l}+2 \left(\sigma ^{k l}\right)_{\alpha }^{\gamma },\nonumber\\
\sigma ^k{}_{\gamma  \dot{\zeta }} \left(\bar{\sigma }^{l m}\right)^{\dot{\zeta }}{}_{\dot{\alpha }}&=&\frac{1}{2} \eta ^{k m} \sigma ^l{}_{\gamma  \dot{\alpha }}-\frac{1}{2} \eta ^{k l} \sigma ^m{}_{\gamma  \dot{\alpha }}-\frac{1}{2} i \eta _{n o} \epsilon ^{o k l m} \sigma ^n{}_{\gamma  \dot{\alpha }},\nonumber\\
\left(\bar{\sigma }^m\right)^{\dot{\beta } \gamma } \left(\bar{\sigma }^{k l}\right)^{\dot{\alpha }}{}_{\dot{\beta }}&=&-\frac{1}{2} \eta ^{l m} \left(\bar{\sigma }^k\right)^{\dot{\alpha } \gamma }+\frac{1}{2} \eta ^{k m} \left(\bar{\sigma }^l\right)^{\dot{\alpha } \gamma }+\frac{1}{2} i \eta _{n o} \epsilon ^{o k l m} \left(\bar{\sigma }^n\right)^{\dot{\alpha } \gamma },\nonumber\\
\left(\bar{\sigma }^m\right)^{\dot{\alpha } \beta } \left(\sigma ^{k l}\right)_{\beta }^{\gamma }&=&\frac{1}{2} \eta ^{l m} \left(\bar{\sigma }^k\right)^{\dot{\alpha } \gamma }-\frac{1}{2} \eta ^{k m} \left(\bar{\sigma }^l\right)^{\dot{\alpha } \gamma }+\frac{1}{2} i \eta _{n o} \epsilon ^{o k l m} \left(\bar{\sigma }^n\right)^{\dot{\alpha } \gamma },\nonumber\\
\sigma ^m{}_{\beta  \dot{\gamma }} \left(\sigma ^{k l}\right)_{\alpha }^{\beta }&=&-\frac{1}{2} \eta ^{l m} \sigma ^k{}_{\alpha  \dot{\gamma }}+\frac{1}{2} \eta ^{k m} \sigma ^l{}_{\alpha  \dot{\gamma }}-\frac{1}{2} i \eta _{n o} \epsilon ^{o k l m} \sigma ^n{}_{\alpha  \dot{\gamma }},\nonumber\\
\left(\bar{\sigma }^{k l}\right)^{\dot{\alpha }}{}_{\dot{\beta }} \left(\bar{\sigma }^{m n}\right)^{\dot{\beta }}{}_{\dot{\alpha }}&=&\frac{1}{2} i \epsilon ^{k l m n}+\frac{1}{2} \eta ^{k n} \eta ^{l m}-\frac{1}{2} \eta ^{k m} \eta ^{l n},\nonumber\\
\left(\sigma ^{k l}\right)_{\alpha }^{\zeta } \left(\sigma ^{m n}\right)_{\zeta }^{\alpha }&=&-\frac{1}{2} i \epsilon ^{k l m n}+\frac{1}{2} \eta ^{k n} \eta ^{l m}-\frac{1}{2} \eta ^{k m} \eta ^{l n},\nonumber\\
\Lambda ^{k l} \Lambda ^{m n} \partial _k\partial _l\partial _mX&=&0,\qquad \Lambda ^{k l} \Lambda ^{m \alpha } \partial _k\partial _l\partial _mX=0,\nonumber\\
\left(\sigma ^{k l}\right)_{\alpha }^{\beta } \partial _k\partial _lX&=&0,\qquad \left(\sigma ^{k l}\right)_{\alpha }^{\beta } \Lambda ^{n \alpha } \Lambda ^o{}_{\beta } \partial _n\partial _oX=0,\nonumber\\
\Lambda ^{k \alpha } \Lambda ^{l \beta } \partial _k\partial _lX&=&-\frac{1}{2} \epsilon ^{\alpha  \beta } \Lambda ^{k l} \partial _k\partial _lX,\qquad \left(\sigma ^{k l}\right)_{\alpha }^{\beta } \Lambda ^{n \alpha } \Lambda ^m{}_{\beta }=-\left(\sigma ^{k l}\right)_{\alpha }^{\beta } \Lambda ^{m \alpha } \Lambda ^n{}_{\beta },\nonumber\\
\Lambda ^{k l} \Lambda ^{m n} \partial _k\partial _mX&=&-\frac{1}{2} \Lambda ^{k m} \Lambda ^{l n} \partial _k\partial _mX,\qquad \Lambda ^{k l} \Lambda ^{m \alpha } \partial _k\partial _mX=-\frac{1}{2} \Lambda ^{k m} \Lambda ^{l \alpha } \partial _k\partial _mX,
\end{eqnarray}
}
where $X$  stands for a superfield.

Moreover, the following relations of $D$ algebra are also used.
{\small
\begin{eqnarray}
\label{D_algebraic_relation}
\left(\bar{D}_{\dot{\beta }} D_{\alpha } \Phi \right)&=&-2 i \sigma ^k{}_{\alpha  \dot{\beta }} \left(\partial _k\Phi \right),\qquad  \left(\bar{D}^2 D_{\alpha } \Phi \right)=0,\nonumber\\
\left(\bar{D}_{\dot{\alpha }} D^2 \Phi \right)&=&-4 i \epsilon ^{\beta  \zeta } \sigma ^k{}_{\zeta  \dot{\alpha }} \left(\partial _k D_{\beta } \Phi \right),\qquad  \left( \bar{D}^2 D^2 \Phi \right)=16 \left(\square \Phi\right), \nonumber\\
\left(D_{\alpha } \bar{D}_{\dot{\beta }} \Phi ^+\right)&=&-2 i \sigma ^k{}_{\alpha  \dot{\beta }} \left(\partial _k\Phi ^+\right),\qquad \left(D^2 \bar{D}_{\dot{\alpha }} \Phi ^+\right)=0,\nonumber\\
\left(D_{\alpha } \bar{D}^2 \Phi ^+\right)&=&4 i \epsilon ^{\dot{\beta } \dot{\zeta }} \sigma ^k{}_{\alpha  \dot{\zeta }} \left(\partial _k \bar{D}_{\dot{\beta }} \Phi ^+\right),\qquad  \left(D^2 \bar{D}^2 \Phi ^+\right)=16 \left(\square \Phi ^+\right),\nonumber\\
D^2 \bar{D}^2 D^2&=&16 \square D^2,\qquad D^2 \bar{D}^2 D_{\alpha }=-4 i \epsilon ^{\dot{\beta } \dot{\zeta }} {p\sigma }_{\alpha  \dot{\zeta }} D^2 \bar{D}_{\dot{\beta }},\nonumber\\
\bar{D}^2 D^2 \bar{D}_{\dot{\alpha }}&=&4 i \epsilon ^{\beta  \zeta } {p\sigma }_{\zeta  \dot{\alpha }} \bar{D}^2 D_{\beta },\qquad \bar{D}^2 D_{\alpha } \bar{D}_{\dot{\beta }}=-2 i {p\sigma }_{\alpha  \dot{\beta }} \bar{D}^2,\nonumber\\
D_{\alpha } \bar{D}^2 D^2&=&4 i \epsilon ^{\dot{\beta } \dot{\zeta }} {p\sigma }_{\alpha  \dot{\zeta }} \bar{D}_{\dot{\beta }} D^2,\qquad D_{\alpha } \bar{D}_{\dot{\beta }} D^2=-2 i {p\sigma }_{\alpha  \dot{\beta }} D^2,\nonumber\\
\bar{D}_{\dot{\alpha }} D^2 \bar{D}^2&=&-4 i \epsilon ^{\beta  \zeta } {p\sigma }_{\zeta  \dot{\alpha }} D_{\beta } \bar{D}^2,\qquad \bar{D}_{\dot{\alpha }} D_{\beta } \bar{D}^2=-2 i {p\sigma }_{\beta  \dot{\alpha }} \bar{D}^2,\nonumber\\
D^2 \bar{D}_{\dot{\alpha }} D_{\beta }&=&-2 i {p\sigma }_{\beta  \dot{\alpha }} D^2,\qquad \bar{D}^2 D^2 \bar{D}^2=16 \square \bar{D}^2,\nonumber\\
D_{\alpha } \bar{D}^2 D_{\beta }&=&\frac{1}{2} \epsilon _{\alpha  \beta } D^2 \bar{D}^2-4 i \epsilon ^{\dot{\zeta } \dot{\iota }} {p\sigma }_{\beta  \dot{\iota }} D_{\alpha } \bar{D}_{\dot{\zeta }},\nonumber\\
\bar{D}_{\dot{\alpha }} D^2 \bar{D}_{\dot{\beta }}&=&-\frac{1}{2} \epsilon _{\dot{\alpha } \dot{\beta }} D^2 \bar{D}^2-4 i \epsilon ^{\zeta  \iota } {p\sigma }_{\iota  \dot{\alpha }} D_{\zeta } \bar{D}_{\dot{\beta }},\nonumber\\
D_{\alpha } \bar{D}_{\dot{\beta }} D_{\zeta }&=&-2 i {p\sigma }_{\zeta  \dot{\beta }} D_{\alpha }-\frac{1}{2} \epsilon _{\alpha  \zeta } D^2 \bar{D}_{\dot{\beta }},\nonumber\\
\bar{D}_{\dot{\alpha }} D_{\beta } \bar{D}_{\dot{\zeta }}&=&\frac{1}{2} \epsilon _{\dot{\alpha } \dot{\zeta }} \bar{D}^2 D_{\beta }-2 i {p\sigma }_{\beta  \dot{\zeta }} \bar{D}_{\dot{\alpha }},
\end{eqnarray}
}
where the identities for $p\Lambda$ and $p\sigma$ read
\begin{eqnarray}
\label{simpliyf_ps}
{p\Lambda }_{\alpha } {p\Lambda }_{\beta }&=&\frac{1}{2} \epsilon _{\alpha  \beta } {p\Lambda }^2,\qquad \epsilon ^{\alpha  \beta } \epsilon ^{\dot{\gamma } \dot{\delta }} {p\sigma }_{\beta  \dot{\delta }}=\overline{{p\sigma }}^{\dot{\gamma } \alpha },\nonumber\\
\epsilon _{\alpha  \beta } \epsilon _{\dot{\gamma } \dot{\delta }} \overline{{p\sigma }}^{\dot{\delta } \beta }&=&{p\sigma }_{\alpha  \dot{\gamma }},\qquad \epsilon _{\dot{\alpha } \dot{\beta }} \overline{{p\sigma }}^{\dot{\beta } \iota } \overline{{p\sigma }}^{\dot{\alpha } \zeta }=\epsilon ^{\zeta  \iota } \square,\nonumber\\
\epsilon _{\zeta  \iota } \overline{{p\sigma }}^{\dot{\beta } \iota } \overline{{p\sigma }}^{\dot{\alpha } \zeta }&=&\epsilon ^{\dot{\alpha } \dot{\beta }} \square,\qquad \epsilon ^{\alpha  \zeta } {p\sigma }_{\alpha  \dot{\beta }} {p\sigma }_{\zeta  \dot{\iota }}=\epsilon _{\dot{\beta } \dot{\iota }} \square,\nonumber\\
\epsilon ^{\dot{\beta } \dot{\iota }} {p\sigma }_{\alpha  \dot{\beta }} {p\sigma }_{\zeta  \dot{\iota }}&=&\epsilon _{\alpha  \zeta } \square,\qquad {p\sigma }_{\beta  \dot{\gamma }} \overline{{p\sigma }}^{\dot{\alpha } \beta }=-\delta _{\dot{\alpha } \dot{\gamma }} \square,\nonumber\\
\overline{{p\sigma }}^{\dot{\alpha } \beta } {p\sigma }_{\beta  \dot{\gamma }}&=&-\delta _{\dot{\alpha } \dot{\gamma }} \square.
\end{eqnarray}

\section{Effective Actions}

\setcounter{equation}{0}
\renewcommand\theequation{B\arabic{equation}}

\subsection{Supersymmetry Invariant Subsets}
\label{result_EF}

We list the $74$ subsets $f_i$'s below, each of them is invariant under the $1/2$ supersymmetry transformation.
{\scriptsize 

}

\subsection{Bases}
\label{susy_base}
We list the 74 bases $B_i$'s below, each of them is invariant under the $1/2$ supersymmetry transformation.
{\scriptsize
%
}

\newpage



\begin{thebibliography}{999}



\bibitem{Seiberg:1999vs} 
  N.~Seiberg and E.~Witten,
  ``String theory and noncommutative geometry,''
  JHEP {\bf 9909}, 032 (1999)
  [arXiv:hep-th/9908142].



\bibitem{Douglas:2001ba} 
  M.R.~Douglas and N.A.~Nekrasov,
  ``Noncommutative field theory,''
  Rev.\ Mod.\ Phys.\  {\bf 73}, 977 (2001)
  [arXiv:hep-th/0106048].
  

\bibitem{Szabo:2001kg} 
  R.J.~Szabo,
  ``Quantum field theory on noncommutative spaces,''
  Phys.\ Rept.\  {\bf 378}, 207 (2003)
  [arXiv:hep-th/0109162].


\bibitem{Gomis:2000zz} 
  J.~Gomis and T.~Mehen,
  ``Space-time noncommutative field theories and unitarity,''
  Nucl.\ Phys.\ B {\bf 591}, 265 (2000)
  [arXiv:hep-th/0005129].


\bibitem{Doplicher:1994tu} 
  S.~Doplicher, K.~Fredenhagen, and J.E.~Roberts,
  ``The Quantum structure of space-time at the Planck scale and quantum fields,''
  Commun.\ Math.\ Phys.\  {\bf 172}, 187 (1995)
  [arXiv:hep-th/0303037].


\bibitem{Bahns:2002vm} 
  D.~Bahns, S.~Doplicher, K.~Fredenhagen, and G.~Piacitelli,
  ``On the unitarity problem in space-time noncommutative theories,''
  Phys.\ Lett.\ B {\bf 533}, 178 (2002)
  [arXiv:hep-th/0201222].







\bibitem{Nilles:1983ge} 
  H.P.~Nilles,
  ``Supersymmetry, supergravity and particle physics,''
  Phys.\ Rept.\  {\bf 110}, 1 (1984).



\bibitem{Wess:1992cp} 
  J.~Wess and J.~Bagger,
  ``Supersymmetry and supergravity,''
  Princeton Universiy Press, USA (1992).

\bibitem{Gates:1983nr} 
  S.J.~Gates, M.T.~Grisaru, M.~Rocek, and W.~Siegel,
  ``Superspace or one thousand and one lessons in supersymmetry,''
  Front.\ Phys.\  {\bf 58}, 1 (1983)
  [arXiv:hep-th/0108200].


\bibitem{Ferrara:2000mm}
S.~Ferrara and M.A.~Lled\'{o},
``Some aspects of deformations of supersymmetric field theories,''
JHEP {\bf 0005}, 008 (2000) 
[arXiv:hep-th/0002084].


\bibitem{Klemm:2001yu} 
  D.~Klemm, S.~Penati, and L.~Tamassia,
  ``Non(anti)commutative superspace,''
  Class.\ Quant.\ Grav.\  {\bf 20}, 2905 (2003)
  [arXiv:hep-th/0104190].




\bibitem{Ooguri:2003qp} 
  H.~Ooguri and C.~Vafa,
  ``The C-deformation of gluino and nonplanar diagrams,''
  Adv.\ Theor.\ Math.\ Phys.\  {\bf 7}, 53 (2003)
  [arXiv:hep-th/0302109].

\bibitem{Ooguri:2003tt} 
  H.~Ooguri and C.~Vafa,
  ``Gravity induced C-deformation,''
  Adv.\ Theor.\ Math.\ Phys.\  {\bf 7}, 405 (2004)
  [arXiv:hep-th/0303063].


\bibitem{Seiberg:2003yz} 
  N.~Seiberg,
  ``Noncommutative superspace, N = 1/2 supersymmetry, field theory and string theory,''
  JHEP {\bf 0306}, 010 (2003)
  [arXiv:hep-th/0305248].

\bibitem{Berkovits:2003kj} 
  N.~Berkovits and N.~Seiberg,
  ``Superstrings in graviphoton background and N=1/2 + 3/2 supersymmetry,''
  JHEP {\bf 0307}, 010 (2003)
  [arXiv:hep-th/0306226].



\bibitem{Ferrara:2003xy} 
  S.~Ferrara, M.A.~Lled\'{o}, and O.~Macia,
  ``Supersymmetry in noncommutative superspaces,''
  JHEP {\bf 0309}, 068 (2003)
  [arXiv:hep-th/0307039].


\bibitem{Britto:2003aj} 
  R.~Britto, B.~Feng, and S.-J.~Rey,
  ``Deformed superspace, N = 1/2 supersymmetry and nonrenormalization theorems,''
  JHEP {\bf 0307}, 067 (2003)
  [arXiv:hep-th/0306215].



\bibitem{Grisaru:2003fd} 
  M.T.~Grisaru, S.~Penati, and A.~Romagnoni,
  ``Two loop renormalization for nonanticommutative N = 1/2 supersymmetric WZ model,''
  JHEP {\bf 0308}, 003 (2003)
  [arXiv:hep-th/0307099].


\bibitem{Britto:2003kg} 
  R.~Britto and B.~Feng,
  ``N=1/2 Wess-Zumino model is renormalizable,''
  Phys.\ Rev.\ Lett.\  {\bf 91}, 201601 (2003)
  [arXiv:hep-th/0307165].


\bibitem{Romagnoni:2003xt} 
  A.~Romagnoni,
  ``Renormalizability of N=1/2 Wess-Zumino model in superspace,''
  JHEP {\bf 0310}, 016 (2003)
  [arXiv:hep-th/0307209].








\bibitem{de Boer:2003dn} 
  J.~de Boer, P.A.~Grassi, and P.~van Nieuwenhuizen,
  ``Noncommutative superspace from string theory,''
  Phys.\ Lett.\ B {\bf 574}, 98 (2003)
  [arXiv:hep-th/0302078].






\bibitem{Majid:1996kd} 
  S.~Majid,
  ``Foundations of quantum group theory,''
  Cambridge Universiy Press, UK (1995).


\bibitem{Miao:2014b} 
Y.-G. Miao and X.-D. Wang, ``All order loop renormalizable Wess-Zumino model on Bosonic-Fermionic
  noncommutative superspace," arXiv:1403.5046 [hep-th].


\bibitem{Berezin:1966} 
F.A.~Berezin, ``The method of second quantization,"
 Academic Press,  New York (1966).





\bibitem{Dimitrijevic:2010yv} 
  M.~Dimitrijevic, B.~Nikolic, and V.~Radovanovic,
  ``(Non)renormalizability of the D-deformed Wess-Zumino model,''
  Phys.\ Rev.\ D {\bf 81}, 105020 (2010)
  [arXiv:1001.2654 [hep-th]].














\bibitem{Araki:2003se} 
  T.~Araki, K.~Ito, and A.~Ohtsuka,
  ``Supersymmetric gauge theories on noncommutative superspace,''
  Phys.\ Lett.\ B {\bf 573}, 209 (2003)
  [arXiv:hep-th/0307076].


\bibitem{Lunin:2003bm} 
  O.~Lunin and S.-J.~Rey,
  ``Renormalizability of non(anti)commutative gauge theories with N = 1/2 supersymmetry,''
  JHEP {\bf 0309}, 045 (2003)
  [arXiv:hep-th/0307275].



\bibitem{Penati:2004gh} 
  S.~Penati and A.~Romagnoni,
  ``Covariant quantization of N = 1/2 SYM theories and supergauge invariance,''
  JHEP {\bf 0502}, 064 (2005)
  [arXiv:hep-th/0412041].




\bibitem{Grisaru:2005we} 
  M.T.~Grisaru, S.~Penati, and A.~Romagnoni,
  ``Non(anti)commutative sym theory: Renormalization in superspace,''
  JHEP {\bf 0602}, 043 (2006)
  [arXiv:hep-th/0510175].






\bibitem{Penati:2009sw} 
  S.~Penati, A.~Romagnoni, and M.~Siani,
  ``A renormalizable N=1/2 SYM theory with interacting matter,''
  JHEP {\bf 0903}, 112 (2009)
  [arXiv:0901.3094 [hep-th]].

\bibitem{Bianchi:2009fn} 
  M.S.~Bianchi, S.~Penati, A.~Romagnoni, and M.~Siani,
  ``Nonanticommutative U(1) SYM theories: Renormalization, fixed points and infrared stability,''
  JHEP {\bf 0907}, 039 (2009)
  [arXiv:0904.3260 [hep-th]].






\end{thebibliography}
\end{document}